\documentclass[twocolumn,showpacs,prb,aps,amsmath,amssymb,superscriptaddress]{revtex4}
\usepackage{subfigure}
\usepackage{amsmath}
\usepackage{graphicx}
\usepackage{rotating}

\def\bbbc{{\mathchoice {\setbox0=\hbox{$\displaystyle\rm C$}\hbox{\hbox
to0pt{\kern0.4\wd0\vrule height0.9\ht0\hss}\box0}}
{\setbox0=\hbox{$\textstyle\rm C$}\hbox{\hbox
to0pt{\kern0.4\wd0\vrule height0.9\ht0\hss}\box0}}
{\setbox0=\hbox{$\scriptstyle\rm C$}\hbox{\hbox
to0pt{\kern0.4\wd0\vrule height0.9\ht0\hss}\box0}}
{\setbox0=\hbox{$\scriptscriptstyle\rm C$}\hbox{\hbox
to0pt{\kern0.4\wd0\vrule height0.9\ht0\hss}\box0}}}}

\newcommand{\beq}{\begin{eqnarray}}
\newcommand{\eeq}{\end{eqnarray}}

\newcommand{\bk}{{\bf k}}

\newcommand{\br}{{\bf r}}

\newcommand{\beqa}{\begin{eqnarray}}
\newcommand{\eeqa}{\end{eqnarray}}

\begin{document}

\title{Bogoliubov angle and visualization of particle-hole mixture in
   superconductors.}
\author{K. Fujita}
\affiliation{LASSP, Department of Physics, Cornell University, Ithaca, NY
14853, USA}

\author{Ilya Grigorenko}
\affiliation{Theoretical Division T-11, Center for Nonlinear
Studies, Center for Integrated Nanotechnologies, Los Alamos National
Laboratory, Los Alamos, New Mexico 87545, USA}

\author{J. Lee}
\affiliation{LASSP, Department of Physics, Cornell University, Ithaca, NY
14853, USA}

\author{M. Wang}
\affiliation{LASSP, Department of Physics, Cornell University, Ithaca, NY
14853, USA}

\author{Jian Xin Zhu}
\affiliation{Theoretical Division,   Los Alamos National Laboratory,
Los Alamos, New Mexico 87545, USA}

\author{J.C. Davis}
\affiliation{LASSP, Department of Physics, Cornell University, Ithaca, NY
14853, USA}
\affiliation{CMPMS Department, Brookhaven National Laboratory, Upton, NY 11973, USA
}

\author{H. Eisaki}
\affiliation{Nanoelectronics Research Institute, AIST, Ibaraki, 305-8568, Japan}

\author{S. Uchida}
\affiliation{Department of Physics, The University of Tokyo, Tokyo, 113-0033, Japan}

\author{Alexander V. Balatsky}
\affiliation{Theoretical Division and Center for Integrated
Nanotechnologies, Los Alamos National Laboratory, Los Alamos, New
Mexico 87545, USA}

\date{Printed \today }

\begin{abstract}
Superconducting excitations ---Bogoliubov quasiparticles ---
 are the quantum mechanical mixture of negatively
charged electron (-e)  and positively charged hole (+e). Depending
on the applied voltage bias in STM one can sample the particle and
hole content of such a superconducting excitation. Recent Scanning
Tunneling Microscope (STM) experiments offer a unique insight into
the inner workings of the superconducting state of superconductors.
We propose a new observable quantity for STM studies that is the
manifestation of the particle-hole dualism of the quasiparticles. We
call it a {\em Bogoliubov angle}. This angle measures  the relative
weight of particle and hole amplitude in the superconducting
(Bogoliubov) quasiparticle. We argue that this quantity can be
measured locally by comparing the ratio of tunneling currents at
positive and negative biases.
   This Bogoliubov angle allows one to measure directly the
   energy and position dependent particle-hole admixture and
   therefore visualize   robustness of  superconducting state
   locally. It may also allow one to measure the particle-hole admixture of
   excitations in normal state above critical temperature and thus may be used to measure
   superconducting correlations in pseudogap state.
\end{abstract}
\pacs{Pacs Numbers: }

\maketitle

\vspace*{-0.4cm}

\columnseprule 0pt

\narrowtext \vspace*{-0.5cm}
\section{Introduction}

The dual particle-wave character of microscopic objects is one of
the most striking phenomena in nature. This  dualism is ubiquitous
in the microworld.  Most notably, the two-slit experiments of Stern
and Gerlach revealed the interference and, hence, the wave nature of
electrons.  In the condensed matter systems, such explicit
visualization of the wave nature of the constituent electrons was
missing until just recently.  The breakthrough came when the
researchers from the IBM labs realized that the best way to
elucidate the electrons {\em inside} a material is to place an
impurity in an otherwise perfect crystal structure. By building
corrals of the impurities on the clean surface, and observing the
generated patters through the scanning tunneling microscope  (STM),
the experimenters were able to demonstrate the laws of the wave
optics using the conduction electron waves.\cite{Crommie, Heller,
coral}

The analog of the conduction electrons in the superconductors are
the quasiparticles.  Unlike electrons, the superconducting
quasiparticles do not carry definite charge.  The same quantum
mechanical dualism is at play when one considers the Bogoliubov
quasiparticles in superconducting state: the quasiparticle is a
coherent combination of an electron and its absence (``hole'').
Particle-hole dualism of quasiparticles  is responsible for a
variety of profound phenomena in superconducting state such as
Andreev reflection, the particle-hole conversion process that is
only possible in superconductor.

\begin{figure}[htb]
\begin{center}
\includegraphics[width=6cm]{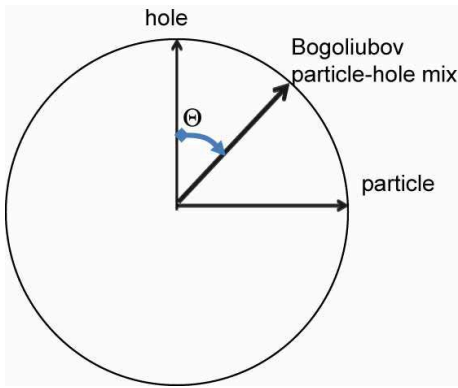}
\caption{ \label{scangle1} Circle parametrizing Bogoliubov
admixture angle is shown. For $\Theta = 0, \pi/2$ the mixture
reduces to purely hole-like and particle-like state. At arbitrary
angle one deals with true Bogoliubov quasiparticles.}
\end{center}
\end{figure}

In this paper we propose a technique to reveal this coherent
particle-hole mixture locally.  In order to discuss the
particle-hole mixture we introduce a quantity that parametrizes the
mixture in terms of an angle, we call this angle a {\em Bogoliubov
angle} (BA), see Fig.~\ref{scangle1}. We argue that STM measurements
allow one to visualize the Bogoliubov angle  maps and thus to reveal
particle hole dualism. Bogoliubov angle  maps as a function of
position and energy offer a tool to investigate strength of
superconducting state locally.

Bogoliubov showed that in order to obtain natural excitations in the
superconducting state one needs to use a linear combination of
particle in hole excitations with  the  coherence factors
$u_{n}(\br_i)$ and
  $v_{n}(\br_i)$. They  describe the unitary transformation from
  particle and hole operators to quasiparticles that are:
  \beqa
  \gamma_{n, \uparrow}(\br_i) = u_n(\br_i)c_{n,
  \uparrow} + v_{n} (\br_i)c^\dag_{n, \downarrow},
  \eeqa
  with the constraints that $\int d \br (|u_{n}(\br_i)|^2+
  |v_{n}(\br_i)|^2)
  = 1$ for any $n$ (normalization) and  $\sum_n ( |u_{n}(\br_i)|^2+
  |v_{n}(\br_i)|^2)
  = 1$ for any $i$ (orthonormality),  $i$ being the site index on our
  lattice.

In the normal state, either $u_{n}(\mathbf{r}_i)$ or
$v_{n}(\mathbf{r}_i)$ are identically zero and there is  no mixing
between the particle- and hole-component of Bogoliubov
quasiparticle. Once  superconductivity sets in, the mixing between
these components develops. This mixing strength can be represented
by
 \beqa
  \Theta_{n}(\br_i) =
  \arctan((\frac{|u_n(\br_i)|^2}{|v_n(\br_i)|^2})^{1/2})
  \label{EQ:SCangle}
  \eeqa
which is a central quantity we are interested in. We define this
quantity as a {\em Bogoliubov angle}. The high resolution STM allows
us to study the spatial dependence of the BA for the states whose
energy can be selected by tuning STM bias.

    Note, we intentionally do not simplify the expression
  in Eq.(\ref{EQ:SCangle}) for the reasons that will be clear in the next section.
 It represents a local mixture between particle and hole
 excitations for an eigenstate $n$ at a given site $i$. For example, for $\Theta_{n}(\br_i) = 0$
 the Bogoliubov excitation will be a hole. In the opposite case of  $\Theta_{n}(\br_i) = \pi/2$ quasiparticle
 is essentially an electron. The angle that corresponds to
  the strongest admixture between particle and hole
 is  $\Theta_{n}(\br_i) = \pi/4 = 45^\circ$. Obviously, in case of inhomogeneous state the BA is a function
 of a position where it is measured and also is a function of  energy $E$. We
  suggest a way to visualize the BA maps that allow us  to develop a more  detailed understanding of the
 superconducting state.
  Previoisly  the alternation of coherence
 factors  $u,v$ as functions of position near impurities has been
 discusses in \cite{Yazdani,MB2000,Hudson,Flatte}. Here we expand this discussion
 by introducing BA. We also will focus more on the spontaneous
 inhomogeneity and not the impurity states that were the focus of
 previous studies.

 The ideas presented here are quite general
 and are applicable to a variety of superconductors, including conventional superconductors.
 Imaging of BA can be performed in any inhomogeneous state.
  One can investigate BA in a variety of states,
 including vortex state and normal state
 with superconducting correlations, e.g. so called pseudogap (PG) state \cite{Yazdani2,Fischer}. To illustrate
 this approach we will use the local STM data obtained on
 high-$T_c$
 superconductor, namely on $Bi_2Sr_2Ca Cu_2 O_{8+\delta}$ material.

The plan of the paper is as follows. We first present a general
theoretical background and define BA from the local tunneling
conductance measurements $dI/dV(\br, V)$  at different bias values
$V$. Then we describe the numerical results for the calculation of
BA.

\section{Theoretical discussion}

\begin{figure}[htb]
\begin{center}
\includegraphics[width=6cm]{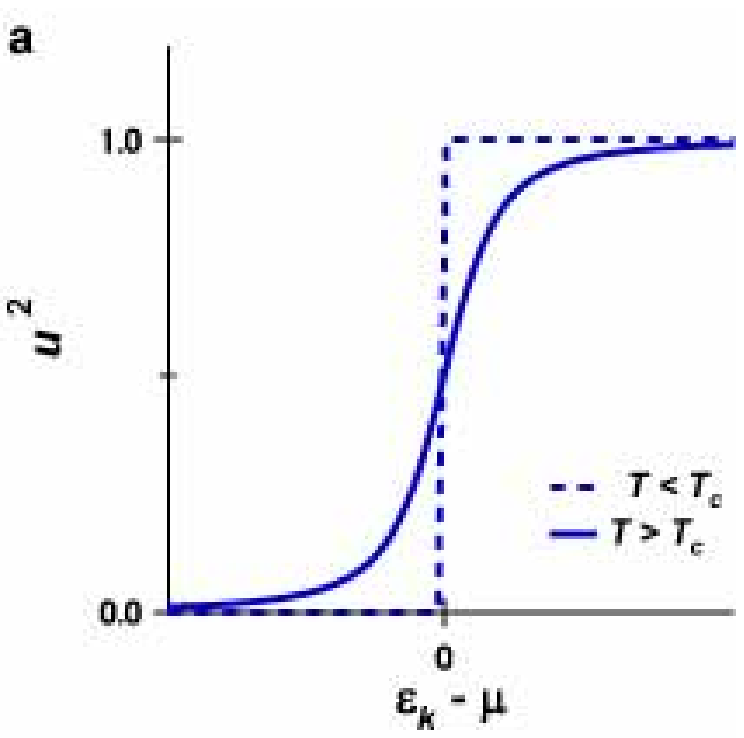}
\includegraphics[width=6cm]{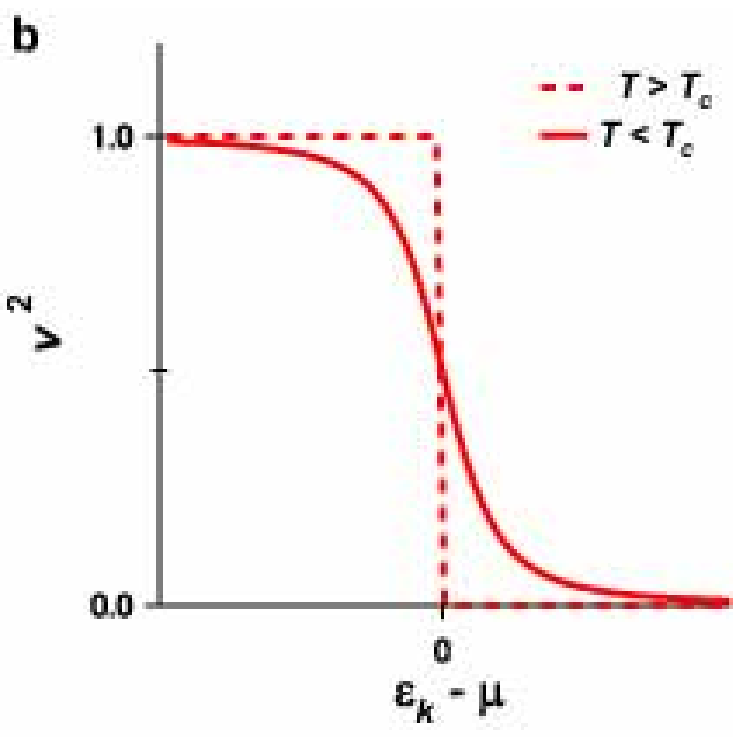}
\includegraphics[width=6cm]{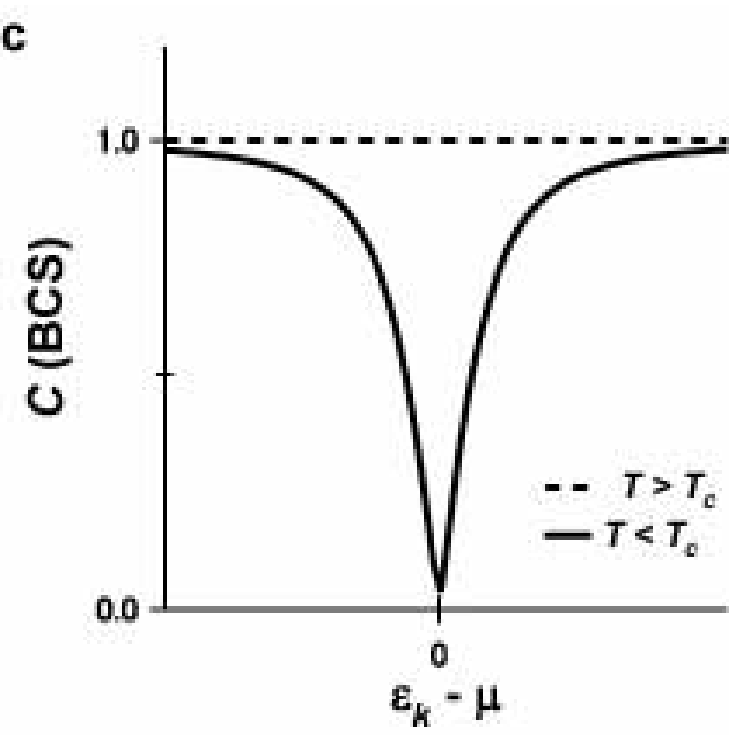}
\caption{\label{concurrence1} (Color online) BCS coherence factors
$u^2(\bk)$ (blue), $v^2(\bk)$ (red) are shown as functions of
energy. The function $C(\bk) = |u^2(\bk)- v^2(\bk)| = |\cos
2\Theta(\bk)|$ (black), shows substantial departures from unity only
in the energy range on the scale of the gap $\Delta$ near the Fermi
energy, where there are substantial pairing correlations.}
\end{center}
\end{figure}

To illustrate the point about BA, we can look at the uniform  BCS
case first. Using Bogoliubov quasiparticles one can introduce BA as:
\beqa \Theta_k  = \arctan [(\frac{|u(\bk)|^2}{|v(\bk)|^2})^{1/2}],
\label{EQ:BCSBA} \eeqa with the conventional coherence factors, see
Fig.~\ref{concurrence1}.

We now turn to inhomogeneous problem where we can no longer use
translational invariance and plane waves as a basis.  It is well
known that
 the  wave function for a BCS superconductor is well captured by a
 following mean field wave function:
 \beqa
 \Psi = \prod_{n}[u_{n}(\br_i - \br_{i'}) +
 v_{n} (\br_{i}- \br_{i'})c^\dag_{n, \uparrow}(\br_i) c^\dag_{n^*,
 \downarrow}(\br_{i'})] |0\rangle,
 \label{EQ:WF1}
 \eeqa
 where $n$ is the eingenvalue index with respective single electron creation
  operators $c^\dag_{n, s}$ in the state $n$ with spin
  $s$. Real space description used here is necessary in case we consider the effects of disorder in single particle potential. Either kind of disorder breaks translational symmetry and the real
  space representation is more natural in this case. The pair wave function is captured in the relative position
  dependence of $u_{n}(\br_i-\br_{i'}), v_{n}(\br_i-\br_{i'})$ components of Bogoliubov
  spinor. The self-consistently defined gap amplitude
  \beqa
  \Delta(\br_i,\br_j) = V_{int} \sum_n
  u_n(\br_i)v_n(\br_j)(1-f(E_n)),
  \label{EQ:Gap1}
  \eeqa
  with $V_{int}$ being interaction, and $f(E_n)$ being the Fermi distribution
   function for a  given quasiparticle excitation spectrum $E_n$.  Here $E_n$ are defined to be positive only.
   We assume that pairing interaction couples only nearest
   neighbors on the lattice $\br_j,\br_i$.

  For a next step in our discussion it is necessary to introduce a
  tunneling conductance as measured by local STM tunneling.
  Introduce   the tunneling conductance  on positive and
  negative bias $E =  \pm  e|V|$ as:
  \beqa
  dI/dV_+(\br_i, E)  &=& F(z, |eV|)\sum_{n,s}|u_{n,s}|^2(\br_i)(-1)f'(E - E_n),\nonumber\\
dI/dV_{-}(\br_i, E)  &=&
F(z,-|eV|)\sum_{n,s}|v_{n,s}|^2(\br_i)(-1)f'(E + E_n), \nonumber \\
\label{EQ:dIdV1} \eeqa where $F(z, \pm|eV|)$ is a function that
measures the matrix elements for tunneling as a function of voltage
bias and tip distance $z$.  Hereafter we assume that it is a smooth
function of energy and at small energies $E \sim 10-100 meV$ it is a
constant.  $f(E)$ is Fermi distribution function. At  very low
temperatures $f'(E)$ becomes a nearly $\delta(E)$ function, a fact
that we will use often. We can simplify the formulas if we introduce
the density of states (DOS) for quasiparticles: $\rho(E) =
\sum_{n,s} \delta(E - E_n)$. For simplicity we will assume
particle-hole symmetry in the normal state.

 Then,  for a given  eigenspectrum and eigenfunctions $u_{n}(\br_i-\br_{i'}),
  v_{n}(\br_i-\br_{i'})$ we can rewrite Eq.(\ref{EQ:dIdV1}) as:
\beqa
  dI/dV_{+}(\br_i, |eV|) &=& -  \int dE \rho( E) |u_{E,s}|^2(\br_i) f'(|eV| -
  E), \nonumber \\ E \geq 0,\nonumber \\
dI/dV_{-}(\br_i, |eV|) &=&  - \int dE \rho( E)
|v_{E,s}|^2(\br_i) f'(|eV| + E), \nonumber \\ E \leq 0. \nonumber \\
  \label{EQ:dIdV2}
  \eeqa
Hence the ratio of $dI/dV$, taken at the same $|E|$, that we label
as $Z(\br_i,|eV|)$,
 will be
 \beqa
  Z(\br_i,|eV| = E) &=& \frac{dI/dV_{+}(\br_i, |eV|)}{dI/dV_{-}(\br_i, |eV|)}
  \nonumber \\
  &=& \frac{|u_{E,s}|^2(\br_i)}{|v_{E,s}|^2(\br_i)} = \tan^2
  \Theta(\br_i,E),
  \label{EQ:ratio1}
  \eeqa
where the last step is taken assuming that there are few, often one
state, that contributes to the summation in Eqs.(\ref{EQ:dIdV1}) a
energy $E = |eV|$. Then Eq.(\ref{EQ:ratio1}) can be inverted as:
\beqa
 \Theta(\br_i,E) =
\arctan[\Huge(\frac{dI/dV_{+}(\br_i, |eV|)}{dI/dV_{-}(\br_i,
|eV|)}\Huge)^{1/2}] \label{EQ:SCangle3} \eeqa this result along with
Eq.(\ref{EQ:SCangle}) is the main result of this section. It allows
a direct determination of Bogoliubov angle $\Theta(\br_i,E)$ from
the experimentally measured tunneling conductances at positive and
negative bias.

BA as a measure of particle-hole admixture appears naturally in the
Anderson mapping \cite{pwa58}  of BCS model on the effective spin
model. We briefly recall the mapping in Appendix A.

To visualize the local quasiparticle states we employ the Scanning
Tunneling Miscroscopy (STM) technique. Crucial aspect of the
electron tunneling into the superconducting state that makes it
qualitatively different from the tunneling in conventional metals is
that  the STM tip contains only the regular electrons which carry a
unit of charge (-e). We can inject either electrons or holes in
superconductor. On the other hand as was pointed out early on
starting with Bogoliubov, quasiparticles that live inside the
superconductor do not possess a well-defined charge. Upon entering
the superconductor, an electron/hole that arrived from the normal
STM tip must undergo a transformation into the Bogoliubov
quasiparticles native to the superconductor \cite{Schrieff}. Hence
electrons that are injected or extracted form superconductor would
need to be ``assembled'' from Bogoliubov excitations. At any  site
and at specific bias this conversion into particles and holes will
depend on relative weights $u_n(\br_i), v_n(\br_i)$. Hence the
intensity of a tunneling signal will depend on these coherence
factors.

Qualitatively, the spatial distribution of tunneling intensity can
be understood as follows.  Respective amplitudes of particle and
hole parts of the Bogoliubov quasiparticle,  are $u_n(\br_i)$ and
$v_n(\br_i)$ for site $i$ and for particular eigenstate $n$.
Consider now a site where, say, $u_n(\br_i)$ is large and close to
1. It follows therefore that for the same site the $v_n(\br_i)$
would have to be small, since the normalization condition is almost
fulfilled by $|u_n(\br_i)|^2$ term alone.
 Similarly, for the
sites where $v_n(\br_i)$ has large magnitude, $u_n(\br_i)$ would
have to be small. Recall now that large $u_n(\br_i)$ component would
mean that quasiparticle has a large {\em electron} component on this
site. Hence the electron  will have large probability to tunnel into
superconductor on this site and the tunneling intensity for
electrons (positive bias) will be large. Conversely, for those sites
the hole amplitude is small $|v(\br_i)| \ll |u(\br_i)|$ and the hole
intensity (negative bias) will be small. Similarly, for sites with
large hole amplitudes $|v(\br_i)| \gg |u(\br_i)|$ the electron
amplitude will be suppressed and this site will be bright on the
hole bias. We observe alternation of the form:
 \beqa
 |v_n(\br_i)|^2 \simeq 1, |u_n(\br_i)|^2 \ll 1,\nonumber\\
 |v_n(\br_i)|^2 \ll 1, |u_n(\br_i)|^2 \simeq 1.
 \label{EQ:uv1}
 \eeqa
 Therefore if there is a particular pattern for the
large particle amplitude (sampled on positive bias) on certain
sites ${i}$, the complimentary pattern of bright sites for hole
tunneling (on negative bias) will develop as a consequence of the
inherent particle-hole mixture in superconductor. This antiphase
behavior is a clear indication of the "natural quasiparticles"
having both particle and hole character. It is the main effect
that can be visualized by considering $\Theta(\br_i,E)$ maps.
Antiphase shift in positive and negative bias intensity is
ubiquitously seen in tunneling spectra.
The "antiphase" behavior of the components $|u_E(\br_i)|^2,
|v_E(\br_i)|^2$ is explained here  as  a case of BA changing from
particle to hole-like configuration on alternating sites. We see
that this is the case  in our numerical simulations, (see Numerical
Simulations below), without any need to assume that only one state
dominates the sum over states in Eq.(\ref{EQ:dIdV2}). So the
phenomenon is more general. We find it easiest to explain assuming
only one term dominating. But given numerical results it holds for
broader cases.

We  discuss it in more details below when we turn to
$\Theta(\br_i,E)$ maps.

\subsection{Particle-Hole asymmetry of normal state}

The question of the underlying band particle-hole asymmetry often
comes up in these materials at low doping. One way to "factor
out" this asymmetry that is extrinsic to the particle-hole mixture
measure, is to factor out the normal state conductances; namely one
can take a ratio of $dI/dV_{\pm}(\br_i, V,T)$ to their proper
normal state values at high temperatures $T>T_c$: \beqa
dI/dV_{\pm}(\br_i,E,T)\rightarrow {\frac{dI/dV_{\pm}(\br_i,
E,T)}{dI/dV_{\pm}(\br_i, E, T> T_c)}}.
 \label{EQ:dIdV3}
  \eeqa
This procedure will factor out the particle hole asymmetry for the
underlying band and will allow  more direct measure of
particle-hole asymmetry.

\section{Imaging Bogoliubov angle in normal and Pseudogap state}

The BA, as defined is not sensitive to the SC quantum phase
fluctuations. Indeed BA is defined as a function of  ratio of the
$|u_E(\br_i )|^2/|v_E(\br_i)|^2$. Therefore $\Theta(\br_i,E)$ can be
defined even in the presence of such phase fluctuations
\cite{Emery}. Thus we propose that the
 discussion about BA may be extended to the normal state.

Imagine we are approaching a normal state of superconductor by
warming it up. We can see that there will be temperature dependence
of the BA. There is no reason to expect an abrupt termination of SC
correlations as one crosses $T_c$. Remnant superconducting
correlations are present above $T_c$\cite{Corson,Xu} and hence one
can still have excitations that will have a particle-hole admixture.
The difference will be that we are no longer in the state with well
defined superconducting Josephson phase.

To illustrate this point consider Bogoliubov-Valatin
transformation in the presence of phase fluctuations: \beqa
\gamma_{n, \uparrow}(\br_i) = u_n(\br_i)c_{n,
  \uparrow} +  \exp(i\phi(\br_i))v_{n} (\br_i)c^\dag_{n,
  \downarrow}.
  \label{EQ:PG1}
  \eeqa
We then can use the same definition for BA, Eq.(\ref{EQ:SCangle})
in this case even in the presence of random Josephson phase
$\phi$. Since one uses amplitudes of $u,v$, spatial phase disorder
does not enter into $\Theta(\br_i,E)$. So for the frozen and
presumably for the slowly varying in time phase fluctuations once
can use the BA as defined and image the local particle-hole
admixture in the normal state.

One would need to take care of thermal broadening of the tunneling
characteristics at higher temperatures. Namely one could divide the
tunneling characteristics by  derivatives of the Fermi thermal
distribution function :

\beqa \Theta_{PG}(\br_i,V)= \arctan [[\frac{dI/dV_{+}(\br_i, V)f'(E
+|eV|)}{dI/dV_{-}(\br_i, V)f'(E-|eV|)}]^{1/2}]. \nonumber \\
\label{EQ:PG2} \eeqa

The problem of dynamic phase fluctuations in the state with
superconducting fluctuations is complicated. More detailed analysis
would require a specific model for the dynamics of the
superconducting phase.
An approach to phase fluctuations in PG state using  localized
Cooper pairs state with no long range phase coherence  was advocated
in \cite{SCZhang}, using STM data \cite{Yazdani2}.

One can also study the behavior  BA for other states, such as flux
phase\cite{Wen} state and density wave states\cite{Chakravarty2000}.
Consider density wave states, e.g. d- density wave state (DDW). DDW
is often mentioned as a possible state that can explain PG
\cite{Chakravarty2000}. In any density wave state, including DDW,
particle-hole symmetry is violated and the poles of single particle
excitations are not appearing  in pairs symmetrically around
chemical potential. Therefore single electron tunneling DOS does not
have the components that appear symmetrically at positive and
negative bias. If there is a particle hole symmetric spectrum for
DDW state it can occur only as a special case at one doping level.

Absence of particle-hole symmetry will be easily detected by BA as
it will tend to pure
 hole or particle  angle, $\Theta\rightarrow 0,\pi/2$. Thus we think
  BA can be used as a spectroscopy tool to detect  presence/absence of
 superconducting correlations in normal state. Another interesting question to address
  is how BA behaves
  upon rising temperature. At low energy it will be close to
  $\pi/2$ but then it can quickly move away to indicate purely
  particle or hole states at $T>T_c$ for non-pairing PG state.



These questions  go beyond the scope of this paper and will be
addressed in separate publication.



\section{Experiment}

In order to visualize the BA, we have performed an experimental
investigation of the Spectroscopic Imaging Scanning Tunneling
Microscopy (SI-STM) measurement on high temperature superconductor
Bi$_2$Sr$_2$CaCu$_2$O$_{8+\delta}$ \cite{Bi2212}. A single crystal
of Bi-2212 grown by floating zone method, is hole-doped by
introducing non-stoichiometric oxygen atoms per unit cell, and its
hole concentration is adjusted for slightly overdoping
($T_{c}=$89K). The crystal is cleaved in the ultra-high vacuum and
immediately inserted into the STM head at $T=$4.2K. To show the BA,
we acquired local density of states (LDOS) images by measuring the
STM tip-sample differential tunneling conductance $g(\vec{r},
V)\equiv dI/dV\mid_{r,V}$ at each location $\vec{r}$ and bias
voltage $V$. Since LDOS$(\vec{r}, E=eV) \propto g(\vec{r}, V)$,
energy and position dependence of the LDOS is obtained.

\begin{figure*}[htb]
\begin{center}
\includegraphics[width=12cm]{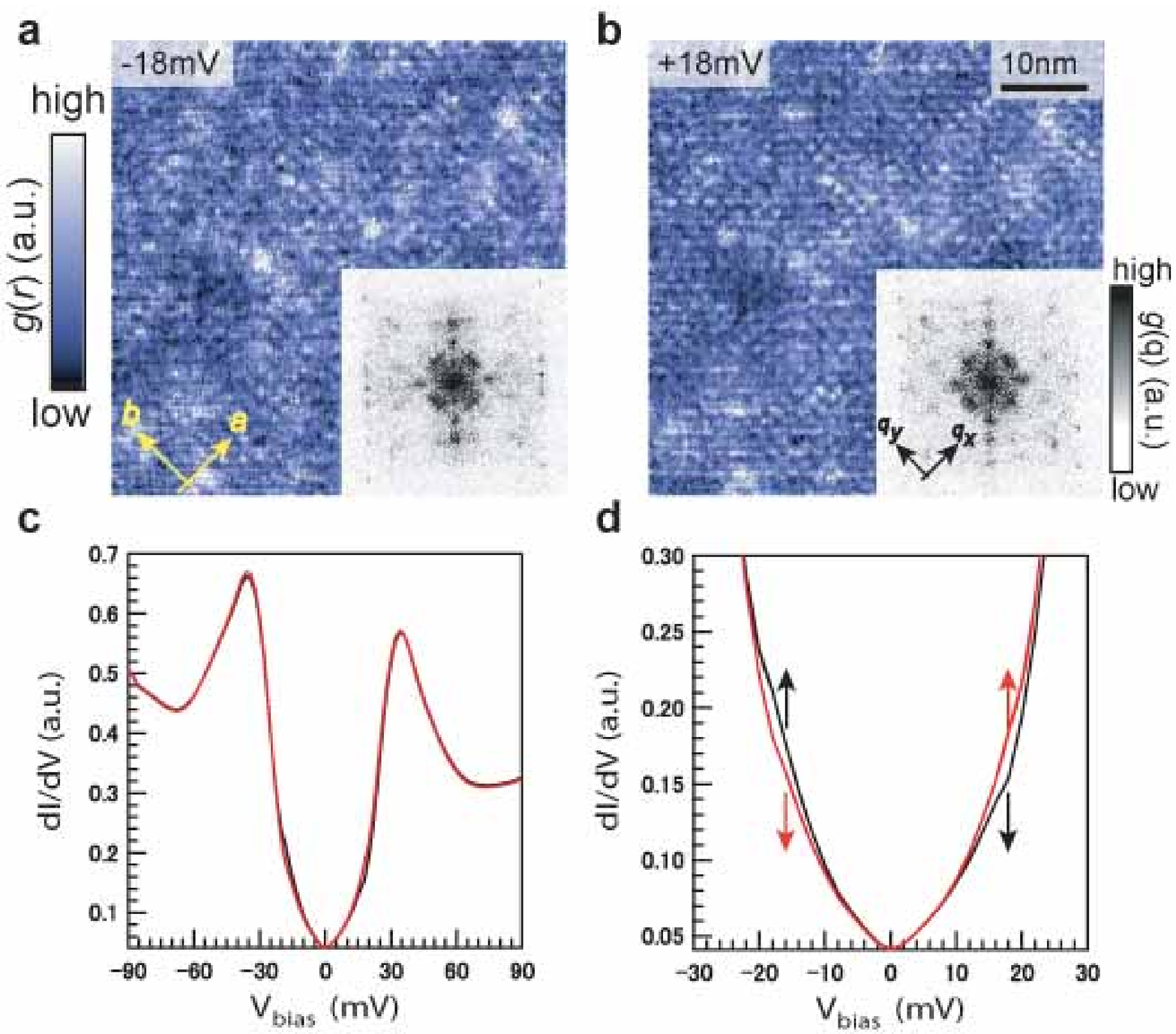}
\caption{\label{fig_exp1} (Color online) $g(\vec{r}, V)$ with 54nm
FOV at $V=$-18mV (\textbf{a}) and $V=$+18mV (\textbf{b}) with their
fourier transforms in the insets. The modulations are visible and
consist of several wave vectors. \textbf{c}, Typical averaged
spectra taken at different area. \textbf{d}, same spectra as
\textbf{c}, but zoomed at the low energy feature below the maximum
gap. Systematic deviation in spectra between the negative and the
positive sample bias is seen, as indicated by arrows.}
\end{center}
\end{figure*}

In Fig.~\ref{fig_exp1}a, we show 54nm $g(\vec{r}, V)$ map at
$V$=-16mV measured on the Bi-2212 surface, showing the spatial
modulations which is interpreted as an interference of the
Bogoliubov quasiparticles \cite{Wang}. A Fourier transform of
$g(\vec{r}, -18\rm{mV})$ in the inset of Fig.~\ref{fig_exp1}
exhibits several fourier spots corresponding to the period of
modulation in real space. These observations are consistent with
previous reports\cite{Kyle, Hoffman}. Although similar modulations
are visible in $g(\vec{r}, +18\rm{mV})$ shown in Fig. \ref{fig_exp1}
which is the same FOV (field-of-view) as Fig.
\ref{fig_exp1}\textbf{a}, one can notice that the spatial phase of
these modulations is different. $dI/dV$-spectra which are averaged
over the regions with the same gap size, where $g(\vec{r},
-18\rm{mV})$ in Fig. \ref{fig_exp1}\textbf{a} is slightly
higher/lower then the average value, see black/red curves in
Fig.~\ref{fig_exp1}\textbf{c}. Overall feature of the spectra taken
with different intensity of $g(\vec{r}, -18\rm{mV})$ with the same
gap are almost identical, however, the significant differences at
low energies in the spectra are seen in the Fig.
\ref{fig_exp1}\textbf{d}. It is obvious that the spectrum with
relatively higher amplitude at the negative sample bias has the
relatively smaller amplitude at the positive sample bias (and vice
versa). This implies that the particle and the hole in the
superconducting state are entangled each other.

\begin{figure*}[htb]
\begin{center}
\includegraphics[width=12cm]{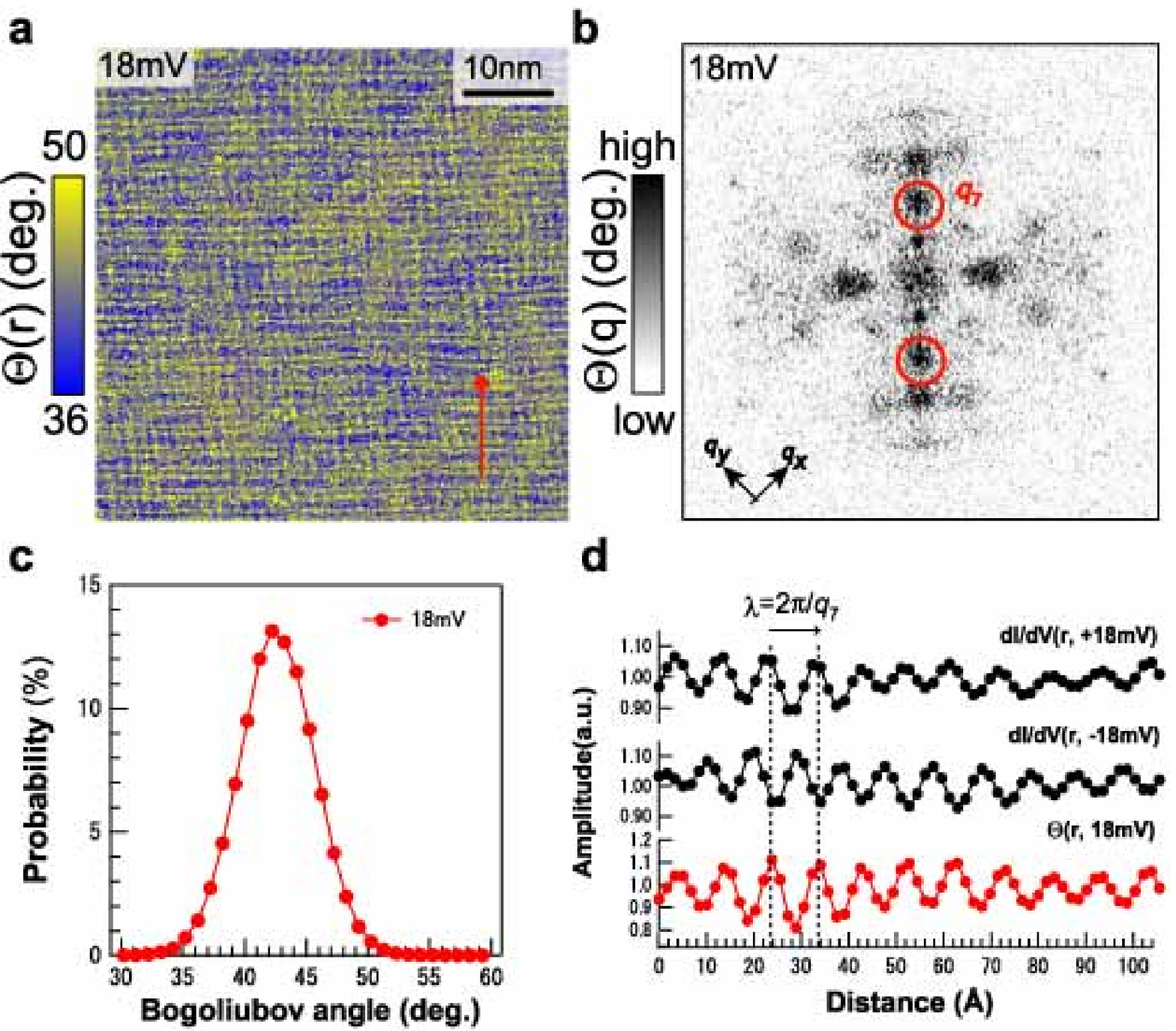}
\caption{\label{fig_exp2}(Color online) \textbf{a},
$\Theta(\textbf{r}_i, V)$ with 54nm FOV at $V=$18mV. \textbf{b},
fourier transform of $\Theta(\textbf{r}_i, V)$ in \textbf{a}.
\textbf{c}, Distribution of $\Theta(\textbf{r}_i, V)$ at $V=$18mV.
\textbf{d}, Spatial evolution of the fourier filtered $dI/dV$ at
18mV and -18mV (black, darker) and $\Theta(\textbf{r}_i,
V=18\textrm{mV})$ (red, lighter) with $2\pi/q_7$ modulation along
the red (light solid) line  starting from the solid (red) circle in
\textbf{a}. }
\end{center}
\end{figure*}

In Fig.~\ref{fig_exp2}a, we calculate a local BA $\Theta
(\textbf{r}_i, V)$ by taking ratio of the positive and the
negative sample bias $g(\vec{r}, V)$, using the following simple
formulas,
\begin{eqnarray}
Z(\textbf{r}_i, V)&\equiv& \frac{\frac{dI}{dV}(\textbf{r}_i, +V)}{\frac{dI}{dV}(\textbf{r}_i, -V)} \label{Z}\\
\Theta (\textbf{r}_i, V)&=&\arctan(\sqrt{Z}). \label{theta}
\end{eqnarray}
Taking ratio has an advantage to cancel out the unknown matrix
element involved in $dI/dV$\cite{Kohsaka}.
Fig.~\ref{fig_exp2}\textbf{a} is the BA map at $V=18$mV with its
fourier transform, and we found that the $\Theta (\textbf{r}_i,
18\textrm{mV})$ shows spatial modulations as well as $dI/dV$ map in
the Fig.~\ref{fig_exp1}, but with more stronger contrast. As seen in
the Fig.~\ref{fig_exp1}\textbf{d}, amplitude of $dI/dV$ between
positive and negative bias are anti-correlated, so that the taking
ratio enhances such structure, namely, spatial modulations. BA map
is essentially different from the $dI/dV$ map, since BA map exhibits
the degree of spatial particle-hole mixture of the Bogoliubov
quasiparticles. However, as evidenced by the fourier transform of
$\Theta (\textbf{r}_i, 18\textrm{mV})$ (
Fig.~\ref{fig_exp2}\textbf{b}), fourier pattern is qualitatively the
same as those of Fig.~\ref{fig_exp1}\textbf{a} and \textbf{b},
indicating that the period of the existing modulation in the BA map
is similar to $dI/dV$ modulations. This similarity supports the
claim that the local particle and hole amplitude are modulated by
scattering. Taking the ratio of $dI/dV$ in Eq.(\ref{Z}) and taking
the BA map in Eq.(\ref{theta}) can therefore be important new tools
to search for the true spatial modulations and individual fourier
spots in the electron density of states.

\begin{figure*}[htb]
\begin{center}
\includegraphics[width=12cm]{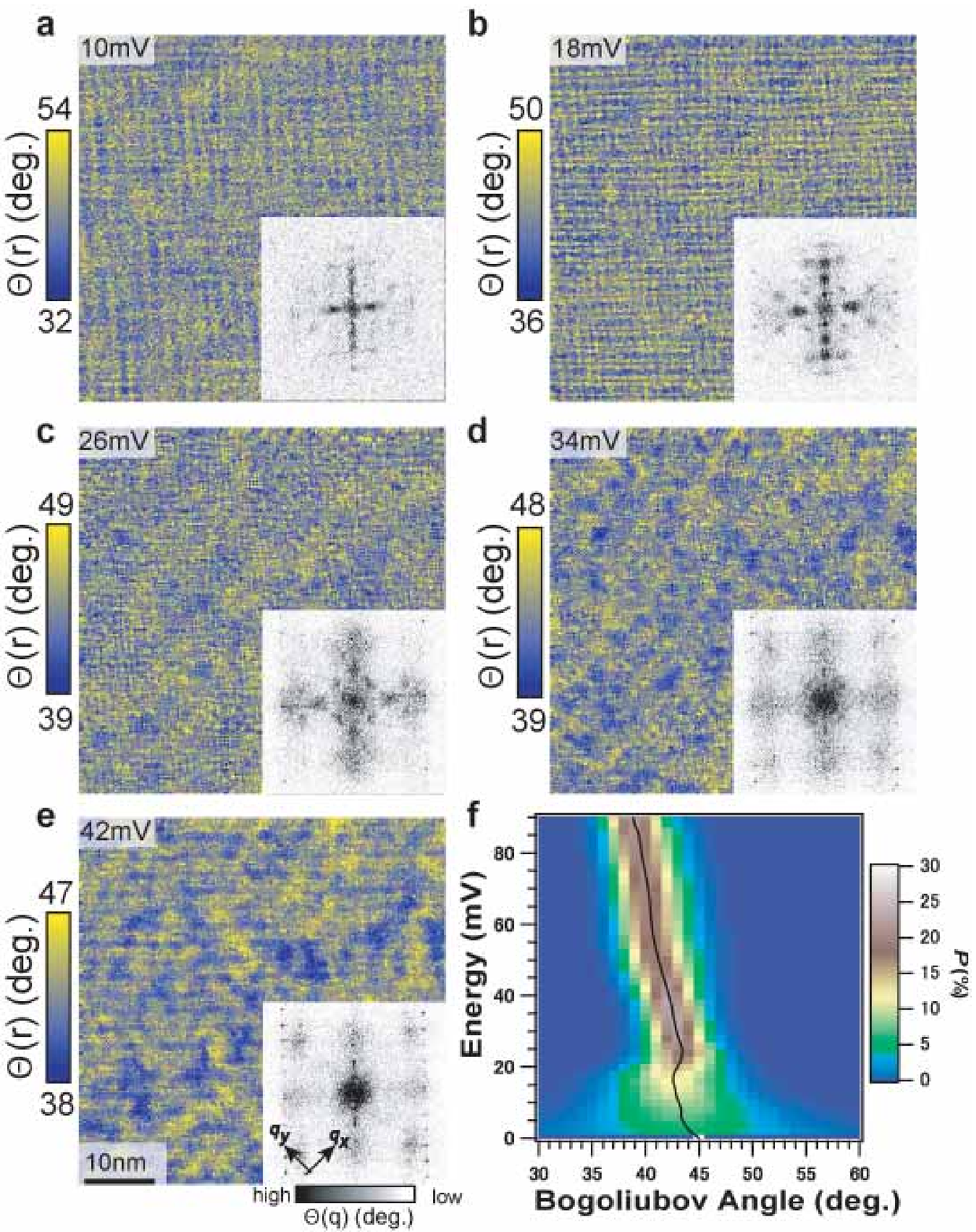}
\caption{\label{fig_exp3}(Color online) \textbf{a, b, c, d, e},
images of the BA at each bias voltages ($V=$10, 18, 26, 34, and
42mV) and their fourier transforms. \textbf{f}, distributions of the
BA at each bias voltages from 0 to 90mV. Peak positions of the
histogram are traced by black line.}
\end{center}
\end{figure*}

In Fig.~\ref{fig_exp2}\textbf{c}, we show the distribution of BA at
 energy, $V = 18 $mV which is peaked at
$\Theta=43^{\circ}$, not exactly at 45$^{\circ}$. One possibility is
that the apparent shift of the distribution is caused by asymmetric
background in the tunneling spectrum \cite{Rantner} that is sampled
more at higher voltage as we shall see in the
Fig.~\ref{fig_exp3}\textbf{f}.
  To visualize the particle- and
hole-like regions more clearly, line cuts of BA at $V=$18mV
 as well as $dI/dV$ at $V=+18$mV and $-18$mV, along the trajectry shown in
Fig.~\ref{fig_exp2}\textbf{a}, are exhibited  in
Fig.~\ref{fig_exp2}\textbf{d}. For simplicity, we only fucus on
the specific $q$-vector in Fig.~\ref{fig_exp2}\textbf{b},
so that the line profiles are taken from the fourier-filtered
$dI/dV$ and $\Theta(\textbf{r}_i, V=18\textrm{mV})$
with $q_7$-vector highlighted by red circle in Fig.~\ref{fig_exp2}\textbf{b}.
Particle- and hole-like regions are
spatially modulating along the line and clearly show the anti-phase behavior
in modulation between $dI/dV$ at +18mV and -18mV.

Figures \ref{fig_exp3}(\textbf{a}-\textbf{e}) show the $\Theta
(\textbf{r}_i, V)$ maps for various bias voltages and their fourier
transforms. With increasing energy, the periods of modulation in
real space change, and corresponding fourier spots in the inset of
the Fig.~\ref{fig_exp3}(\textbf{a}-\textbf{e}) move, following the
octet model \cite{Wang}, and these observations are consistent with
previous reports \cite{Kyle, Hoffman}. In addition to the period of
modulation, one can immediately notice that the pattern of the
spatial modulation changes. At low energies
(Fig.~\ref{fig_exp3}\textbf{a} and \textbf{b}), spatial modulations
are visible all over the field of view. On the other hand, at
$V=$34mV (or $V>$34mV), such modulations tend to be visible in the
restricted area. This difference implies that the different type of
scattering might kick in at $V=$34mV (or $V>$34mV).

In Fig.~\ref{fig_exp3}\textbf{f}, we show the 2D distribution of the
BA in which distributions are normalized at each energies. The
spatial change in the BA map seems to occur as a crossover, and it
can be realized by deviation of the BA from $\Theta=45^{\circ}$ in
the Fig.~\ref{fig_exp3}\textbf{f}. The energy which differentiates
the spatially coherent excitations and the localized excitations is
estimated as $\sim$26mV (less than mean $\Delta \sim $40mV) where BA
starts to monotonically decrease.

The visualization of the BA shed light to understand the
quasiparticle excitations in the superconducting state. The
interferences of the Bogoliubov quasiparticles can be understood
as a spatial variation of relative weight of the particle and the
hole amplitude which is represented by the BA. The BA can be a
measure of the energy scale of the coherent excitations which
split the type of the modulation structure in real space. And,
since the spatial modulations in the electronic structure
are revealed much more clearly in the BA map,
this provides an excellent new technique
 to determine the momentum
space ($q$-space) electronic structure using SI-STM. Hanaguri
\textit{et al.} have recently demonstrated the power of this
technique with the discovery of the interference of the Bogoliubov
quasiparticles in Na-CCOC \cite{Hanaguri}.

\section{Numerical simulations }

 We implement simple but
realistic model of the optimally  doped cuprate superconductor with
disorder. We use a simple BCS solution to illustrate the approach on
how one can visualize the supercondcuting admixture of particles and
holes in the natural Bogoliubov excitations in superconducting
state. Even though the model is simplistic the approach itself is
quite general.

To model the high-temperature superconductors we utilize the
highly-anisotropic structure of the cuprates and focus on a single
layer of the material.  In the simplified model, the conduction
electrons live on the copper sites, $i$, and can hop to the
neighboring sites, $j$, with a certain probability measured by the
quantity $t$.  In addition to that, the electrons that occupy the
neighboring sites feel mutual attraction of a strength $V_{int}$.
Formally, this model is represented by the Hamiltonian, \beq
\label{1}H_0 = -t \sum_{<i,j>, \sigma} {c^{\dagger}_{i\sigma}
c_{j\sigma}}
    -V_{int} \sum_{<i,j>}{n_i n_j},
\eeq were a quantum-mechanical operator $c^{\dagger}_{i\sigma}$
creates an electron on site $i$, the operator $c_{j\sigma}$
eliminates an electron from the site $j$, and $n_i =
c^{\dagger}_{i\uparrow}c_{i\uparrow} +
c^{\dagger}_{i\downarrow}c_{i\downarrow}$
 represents the electron density on site $i$.  The
electron spin, $\sigma$, can point up or down.  This model, referred
to  as the $t-V$ model, is known to produce the d-wave pairing for
the electron densities close to one electron per lattice site, and
has been successfully used to describe  strong impurities in a
d-wave superconductor \cite{balatsky}. The local impurity is
introduced by modifying the electron energy on a particular site.
The corresponding correction to the Hamiltonian is
\beq\label{eq:Himp} H_{imp} = V^{imp}(n_{i\uparrow} +
n_{i\downarrow}). \eeq This term is the potential part of the
impurity energy that couples to the total electronic density on site
$i$. We solve the impurity problem in the Hartree-Fock
approximation, which replaces the two-body interaction in $H_0$ with
an effective singe-electron potential. Our goal is to use $V_{imp}$
to investigate the spatial distribution of BA as a function of
position and energy.

\begin{figure}[htb]
\begin{center}
\includegraphics[width=7cm]{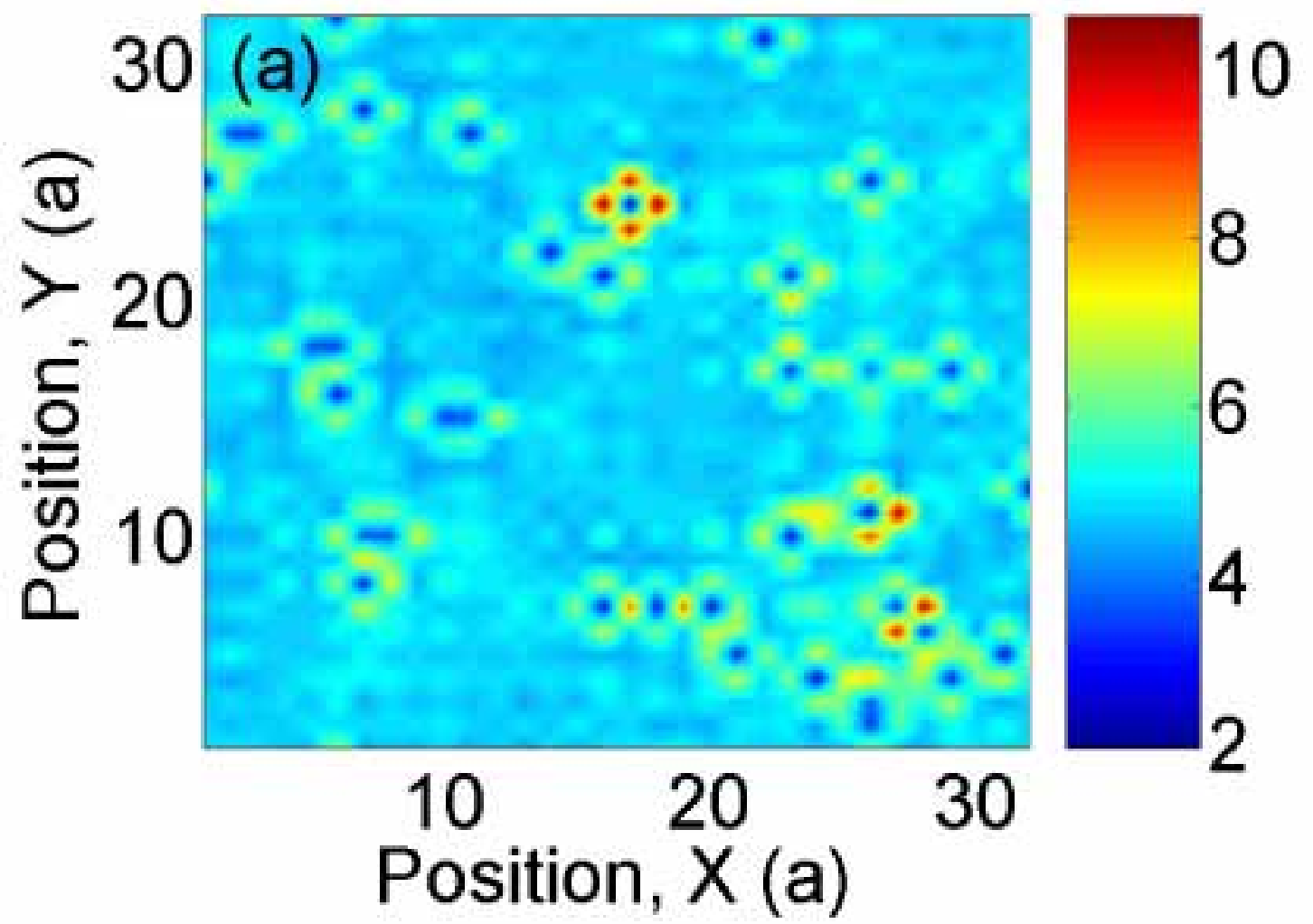}
\includegraphics[width=7cm]{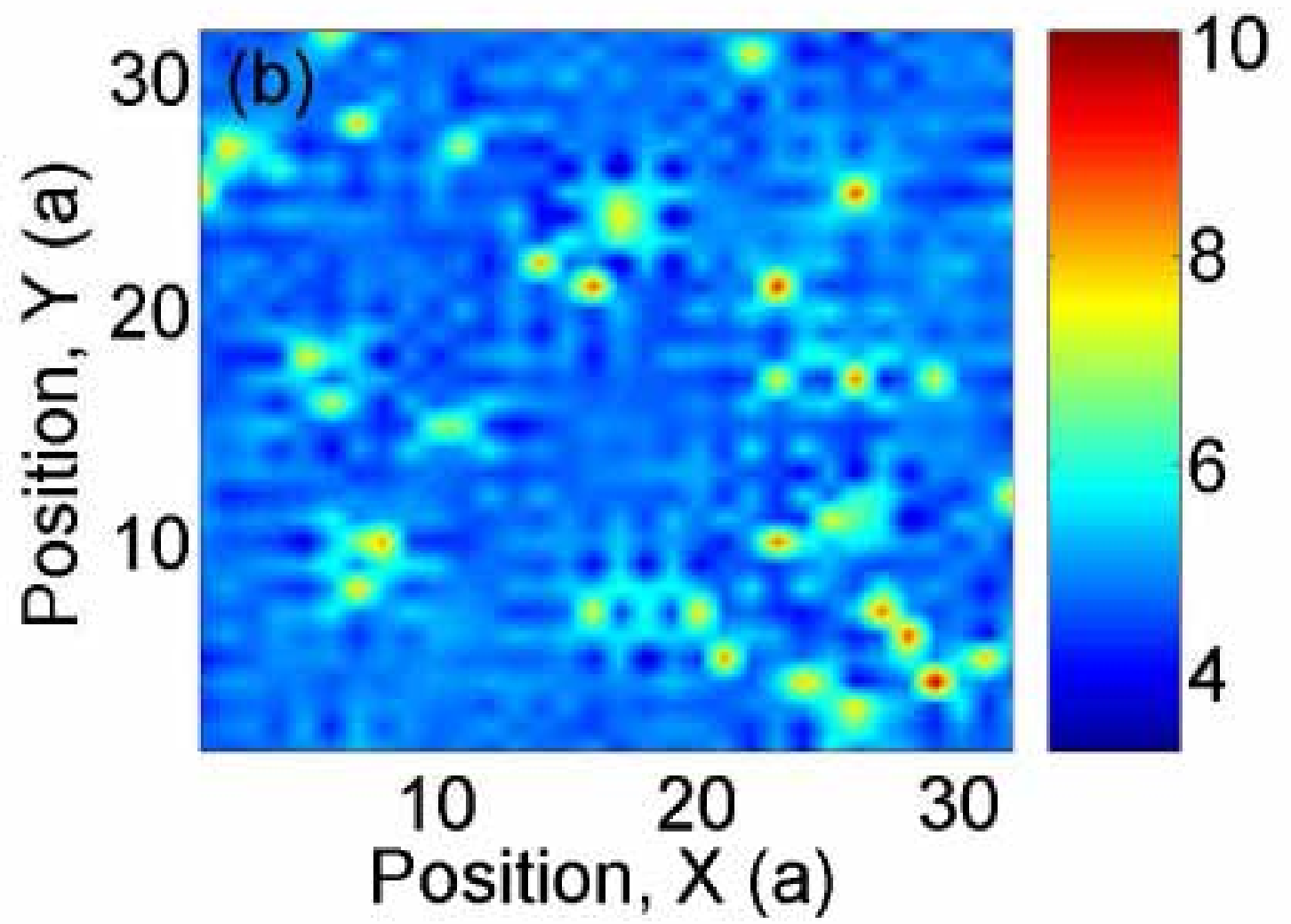}
\includegraphics[width=7cm]{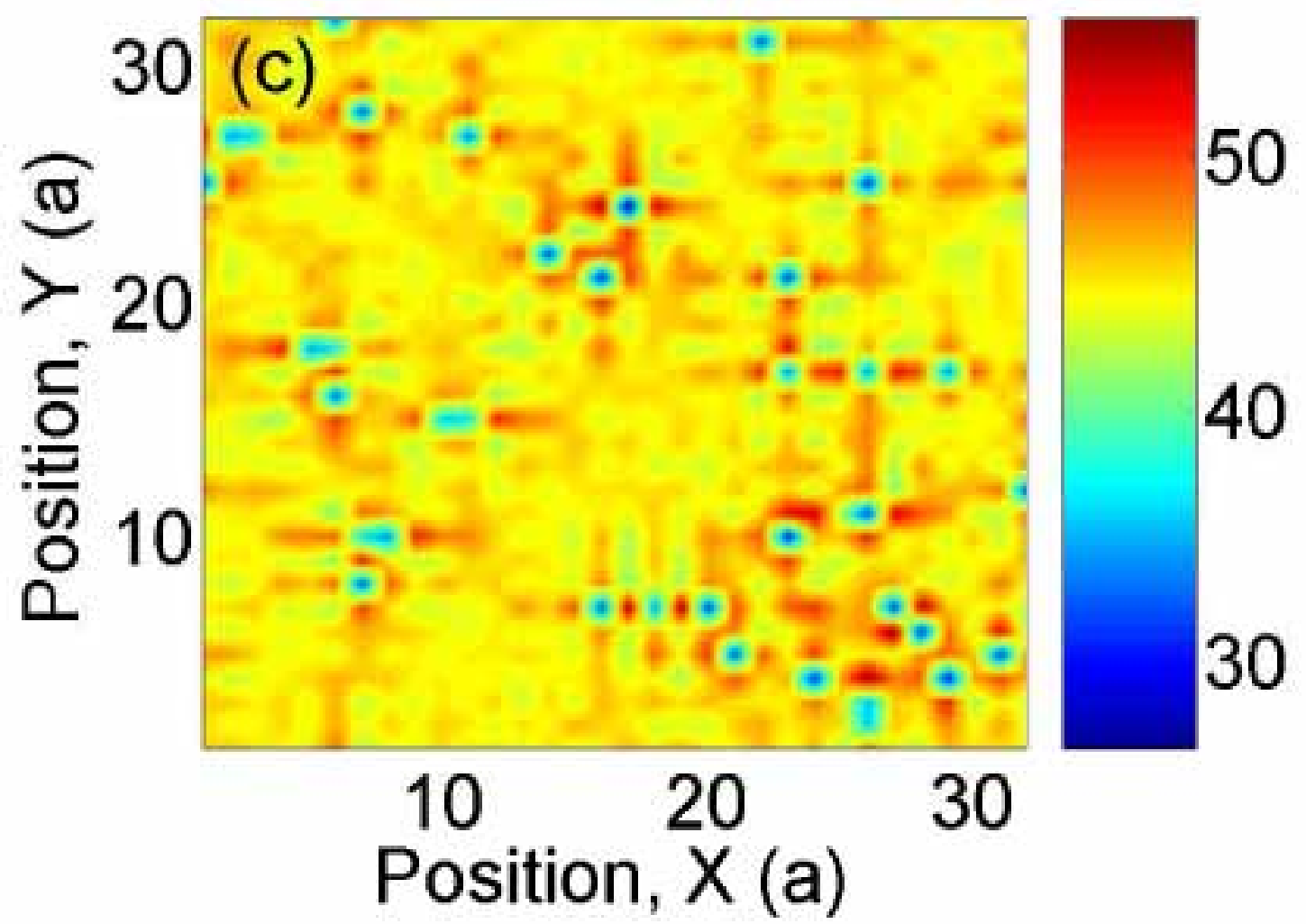}
\includegraphics[width=7cm]{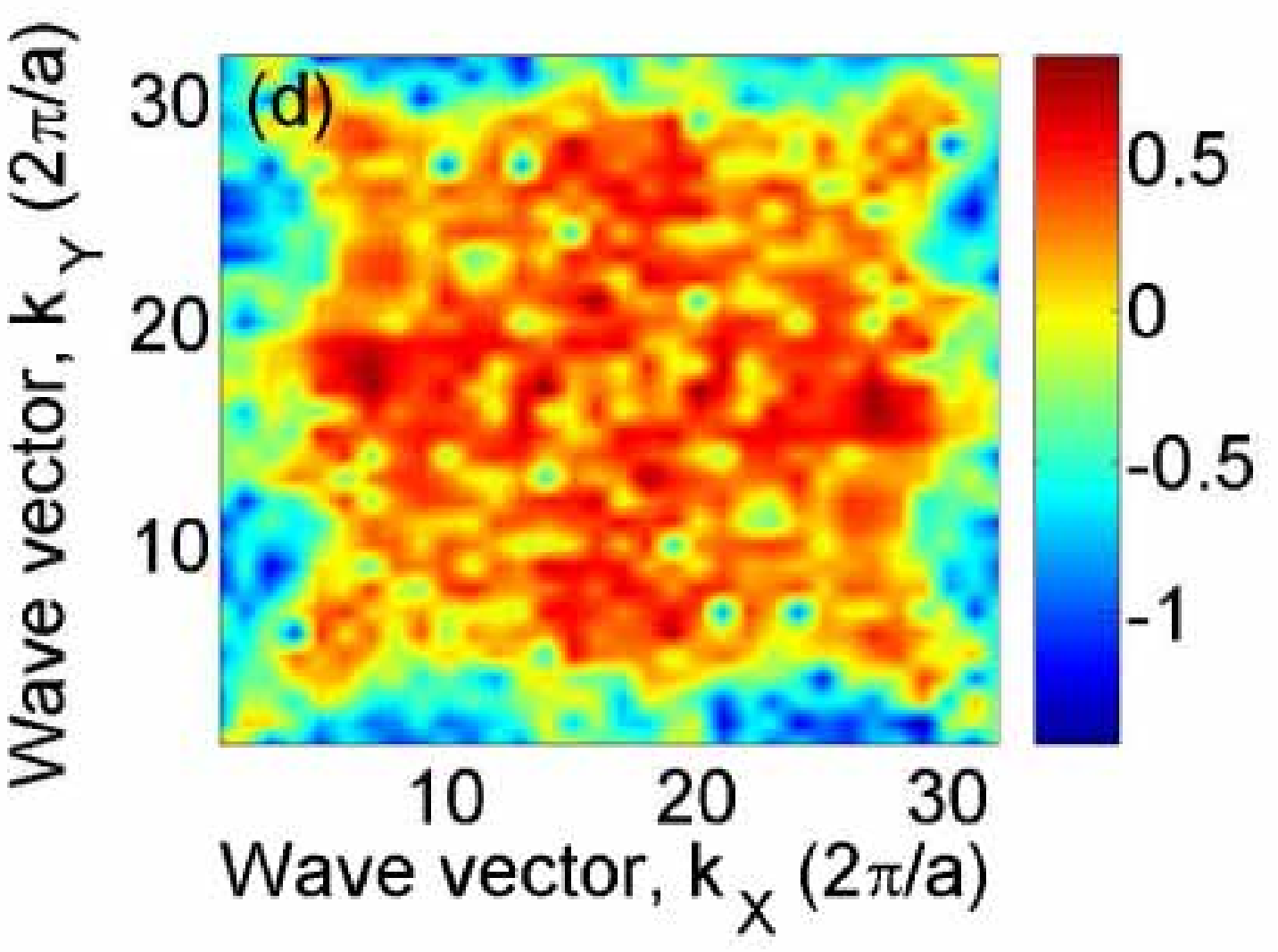}
\caption{\label{fig_theor1} (Color online) \textbf{a,b,c,d},
calculated LDOS on a square $32\times32$ lattice at $T=0$. We assume
$40$ randomly placed impurities with individual impurity strength
$V^{imp}=1 t$. The pairing strength is set $V_{int}=-2 t$ and
chemical potential $\mu=0$ (see text). \textbf{a}, calculated local
$\frac{dI}{dV}$ tunneling conductance at positive bias $V=0.4 t$.
\textbf{b}, calculated local $\frac{dI}{dV}$ tunneling conductance
at negative bias $V=-0.4 t$. \textbf{c},  corresponding Bogoliubov
angle $\Theta(x,y)$. \textbf{d}, logarithm of the absolute value of
Fourier transform of BA $\log_{10}(|\Theta(k_x,k_y)|)$ with
subtracted average value $<\Theta(x,y)>=45^\circ$.}
\end{center}
\end{figure}

\begin{figure}[htb]
\begin{center}
\includegraphics[width=7cm]{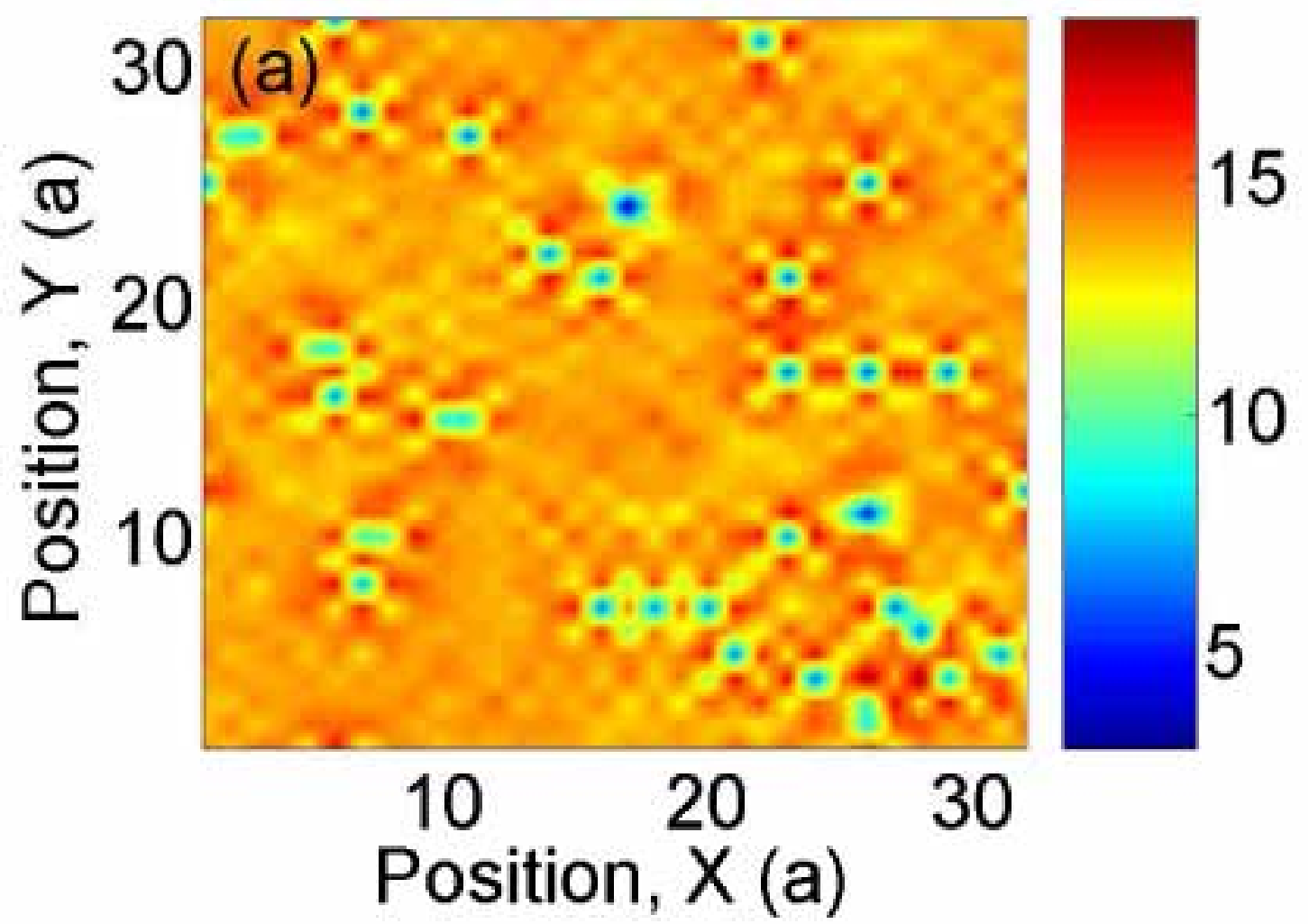}
\includegraphics[width=7cm]{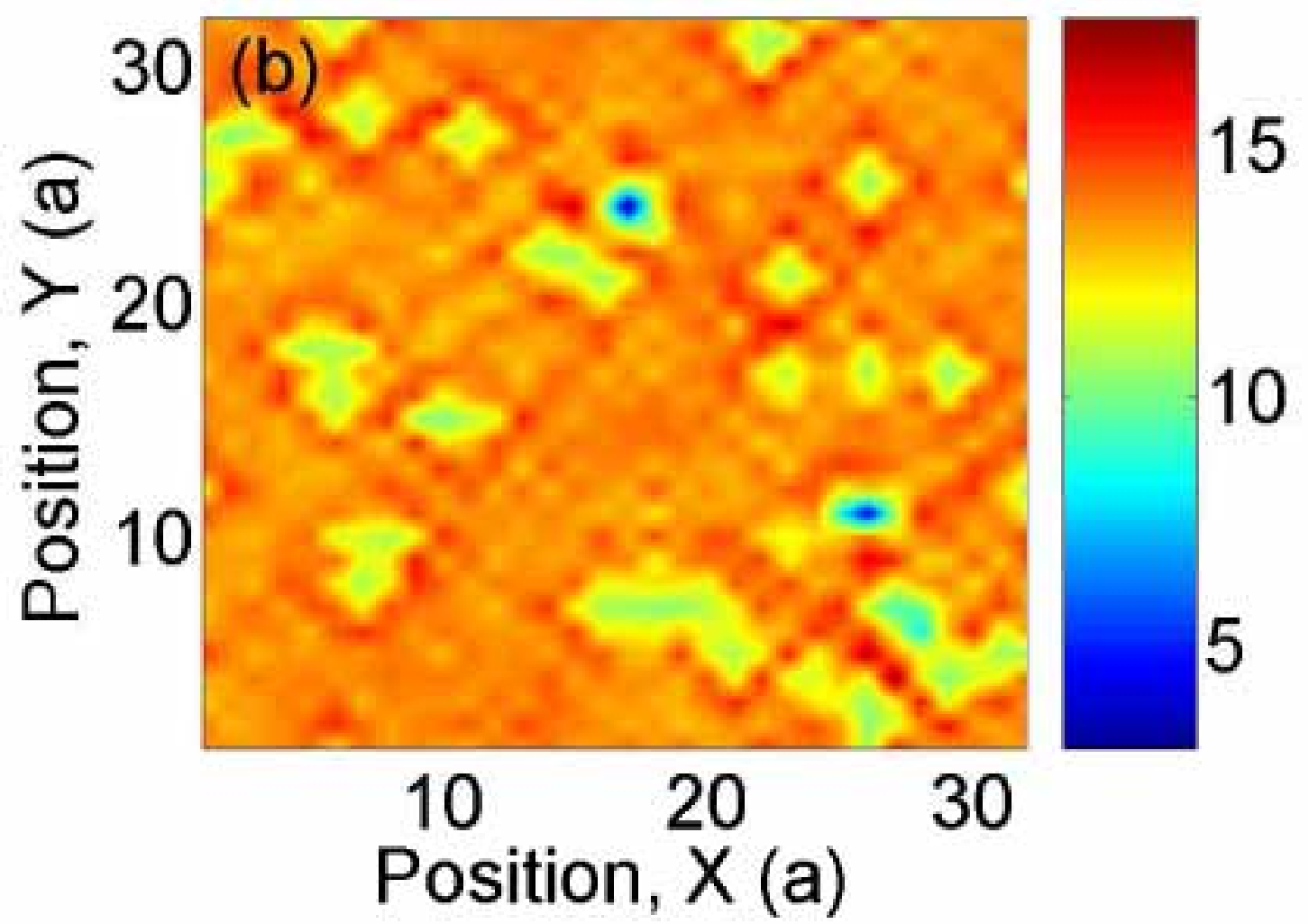}
\includegraphics[width=7cm]{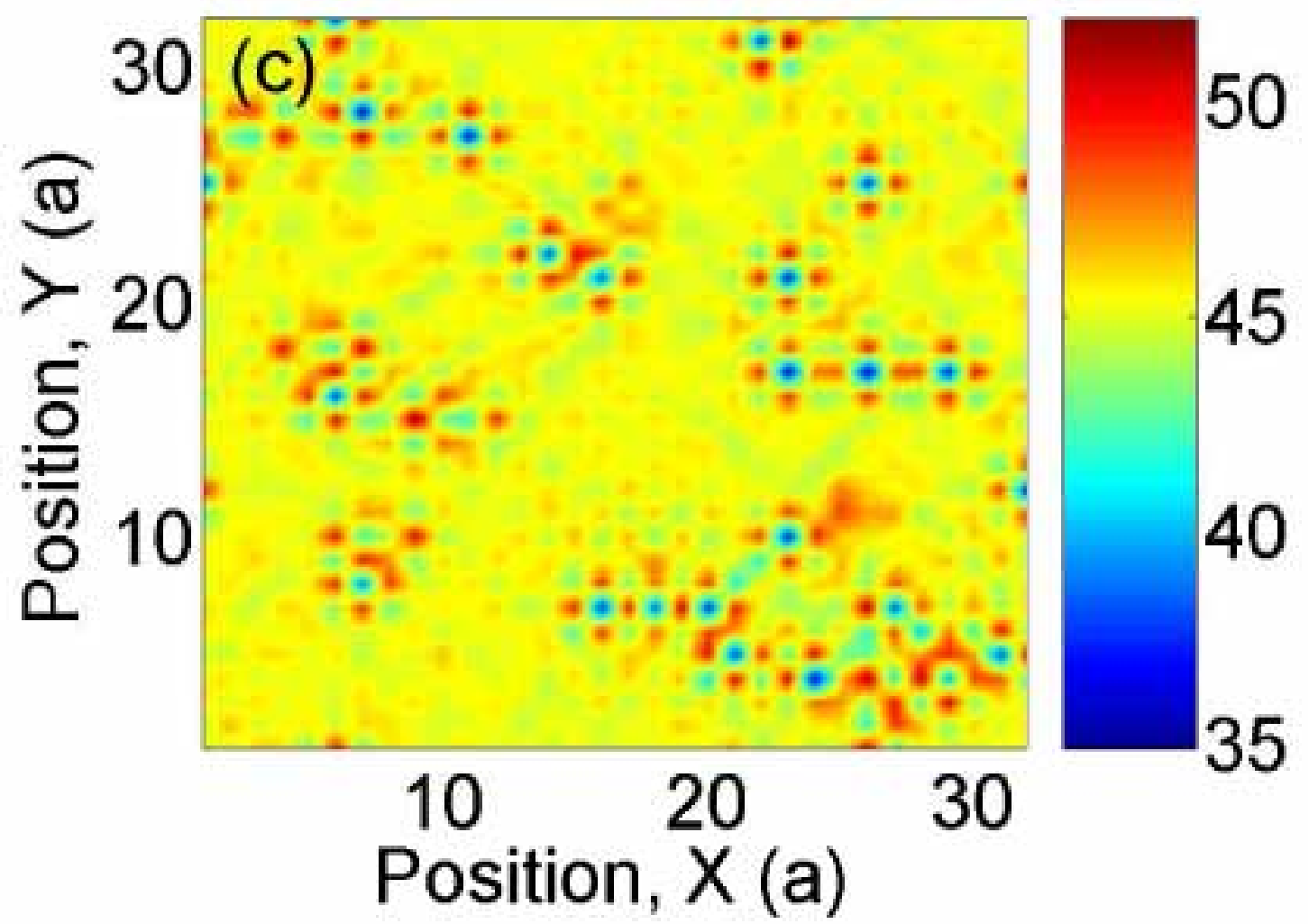}
\includegraphics[width=7cm]{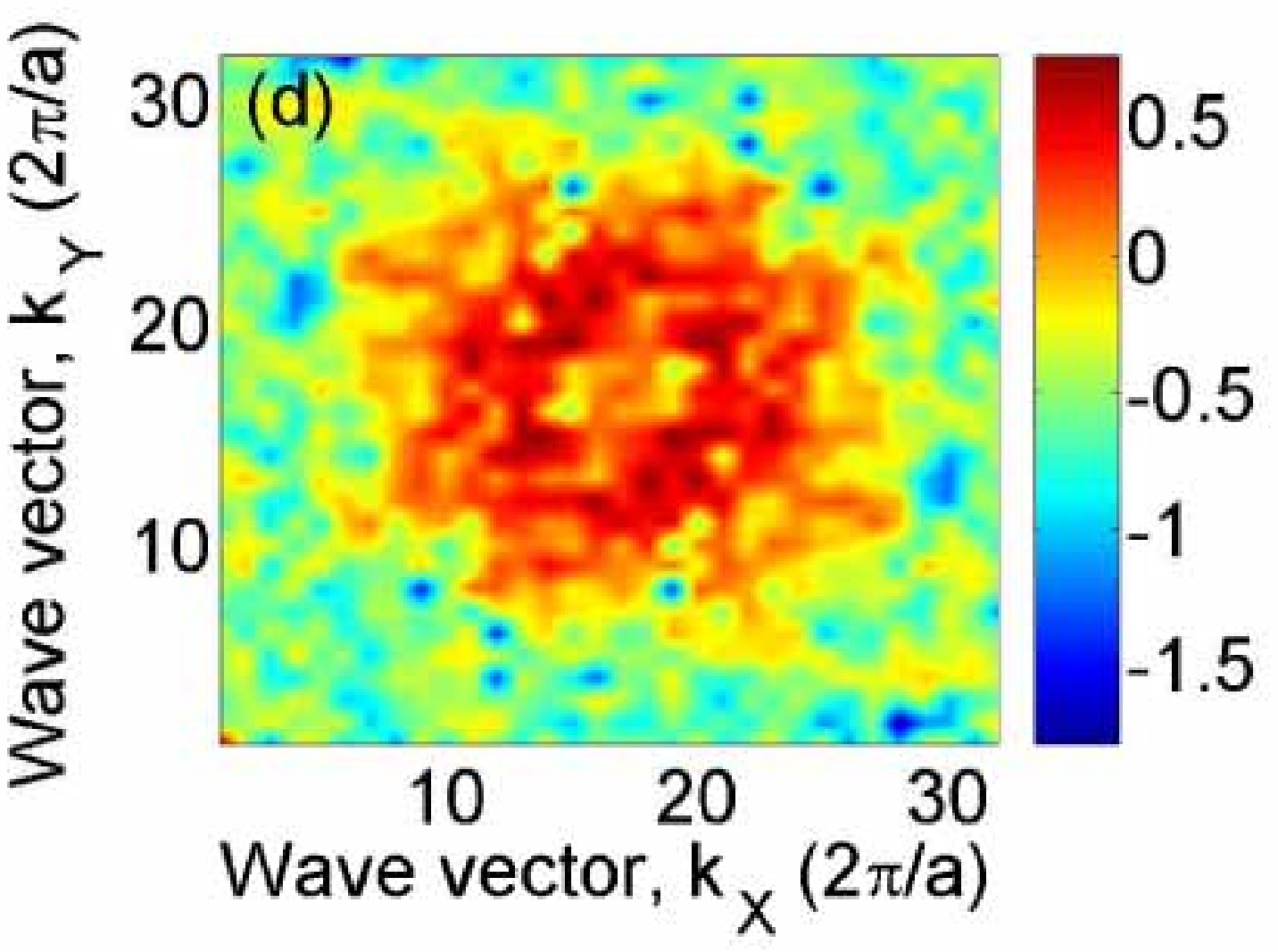}
\caption{\label{fig_theor2} (Color online) \textbf{a,b,c,d},
calculated LDOS on a square $32\times32$ lattice at $T=0$. We assume
$40$ randomly placed impurities with individual impurity strength
$V^{imp}=1 t$. The pairing strength is set $V_{int}=-2 t$ and
chemical potential $\mu=0$ (see text). \textbf{a}, calculated local
$\frac{dI}{dV}$ tunneling conductance at positive bias $V=0.8 t$.
\textbf{b}, calculated local $\frac{dI}{dV}$ tunneling conductance
at negative bias $V=-0.8 t$. \textbf{c},  corresponding Bogoliubov
angle $\Theta(x,y)$. \textbf{d}, logarithm of the absolute value of
Fourier transform of BA $\log_{10}(|\Theta(k_x,k_y)|)$ with
subtracted average value $<\Theta(x,y)>=45^\circ$.}
\end{center}
\end{figure}

\begin{figure}[htb]
\begin{center}
\includegraphics[width=7cm]{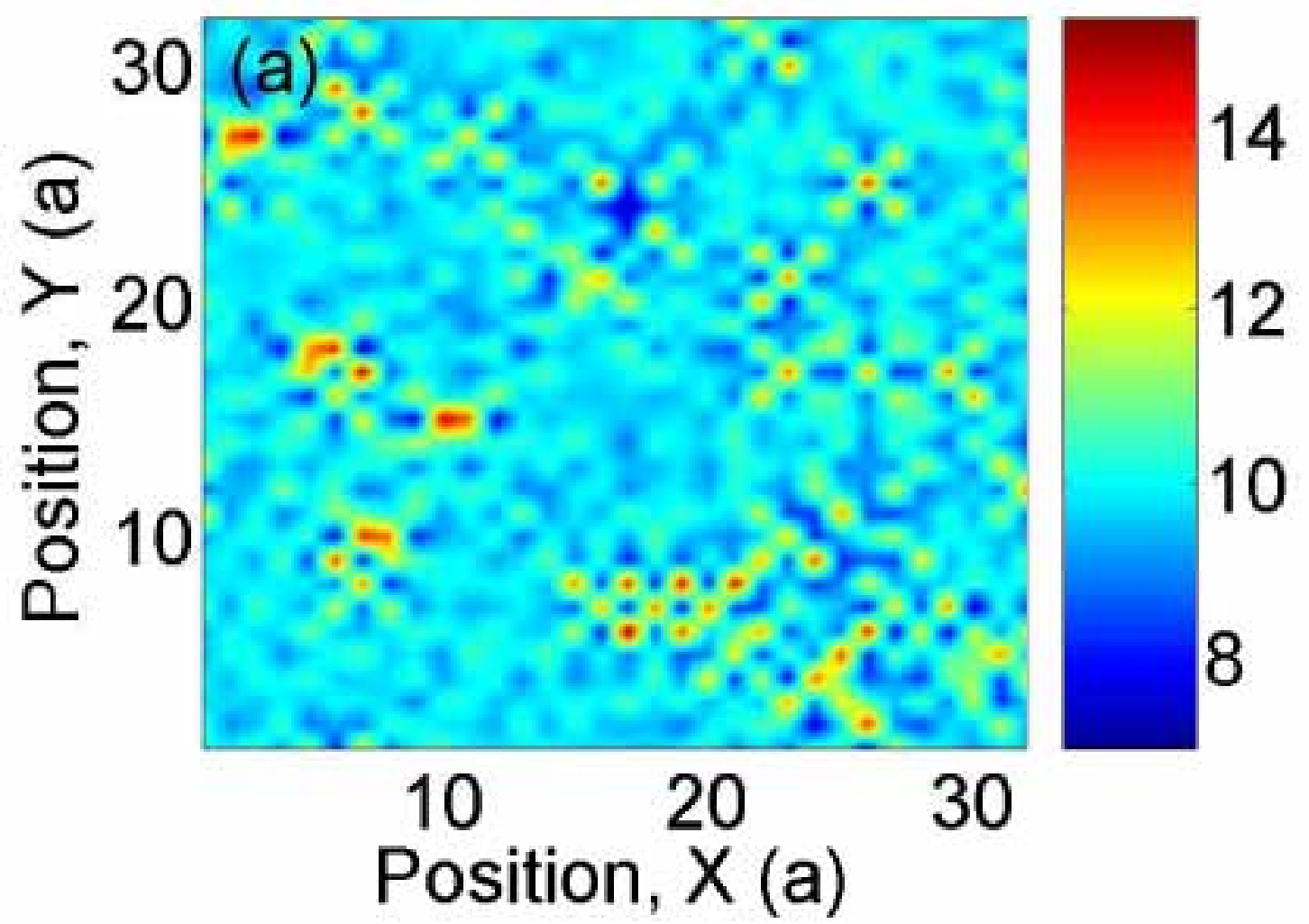}
\includegraphics[width=7cm]{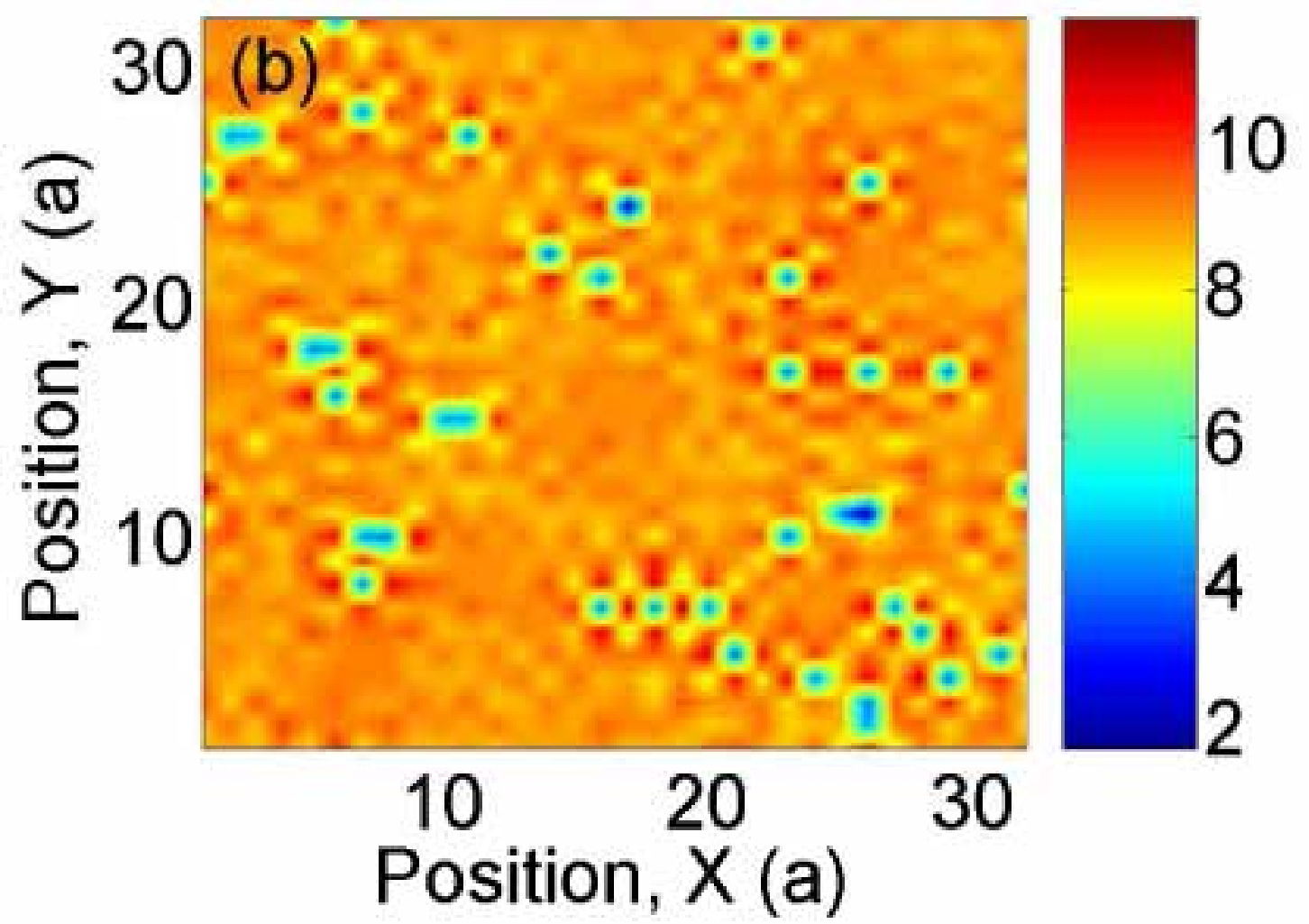}
\includegraphics[width=7cm]{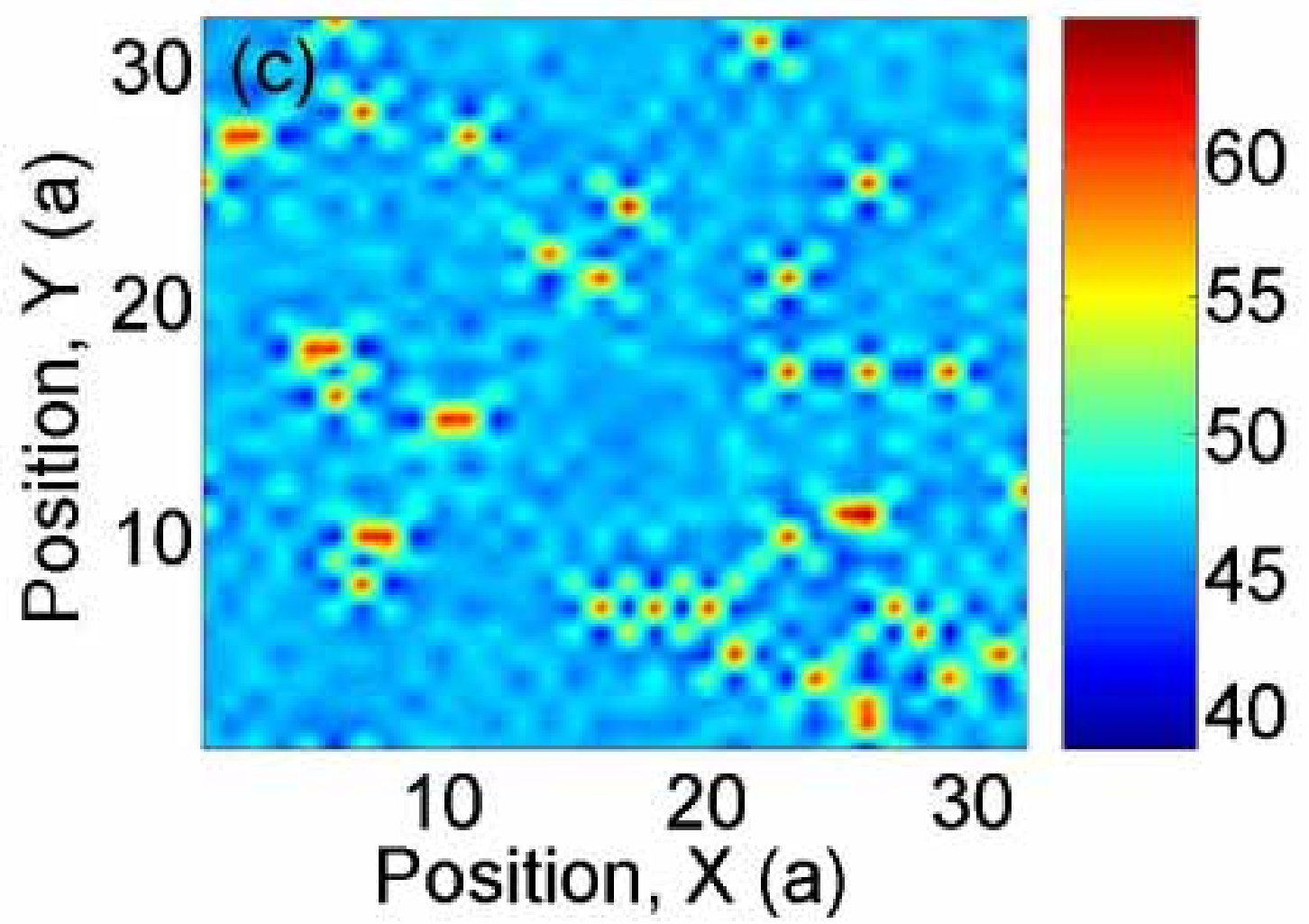}
\includegraphics[width=7cm]{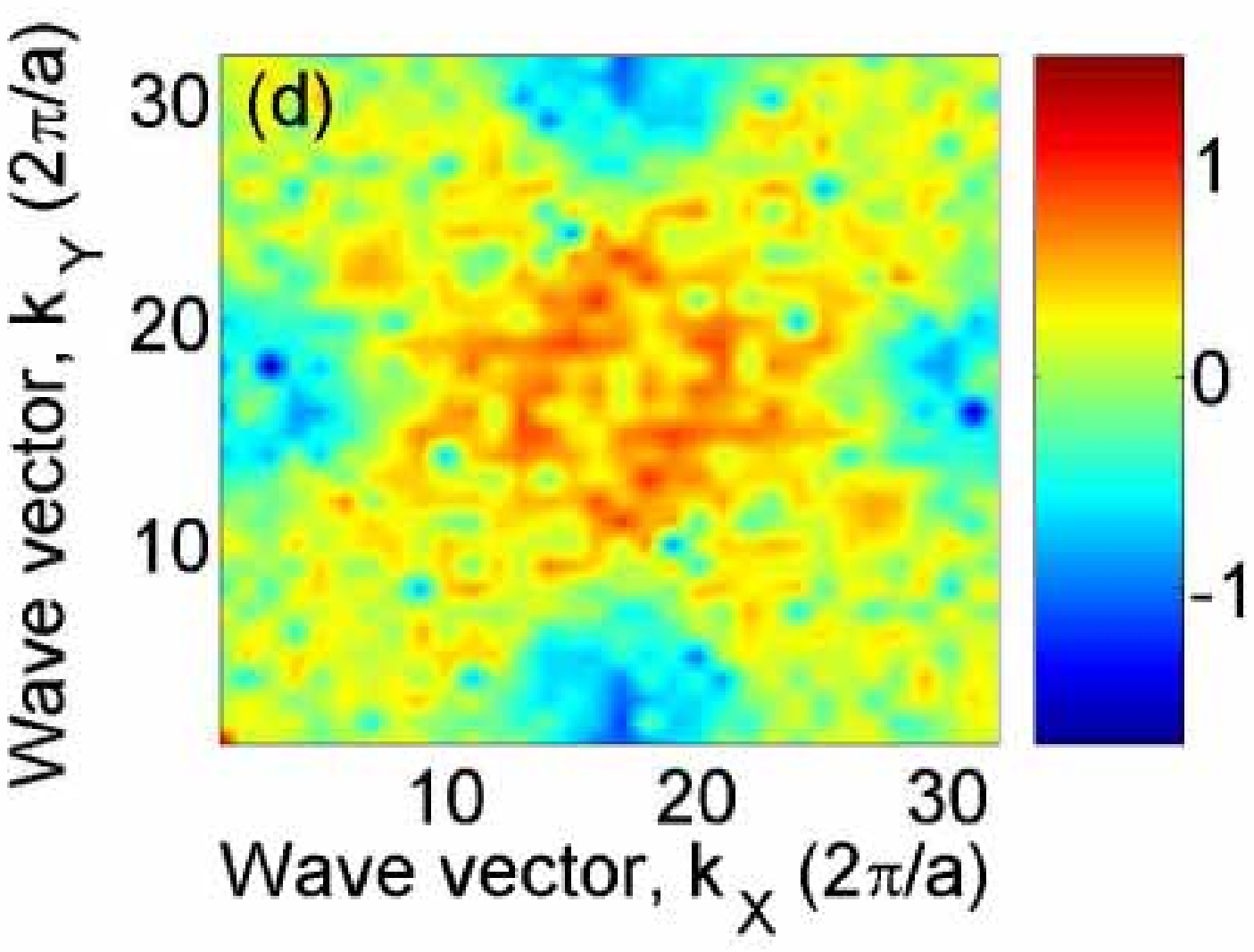}
\caption{\label{fig_theor3} (Color online) \textbf{a,b,c,d},
calculated LDOS on a square $32\times32$ lattice at $T=0$. We assume
$40$ randomly placed impurities with individual impurity strength
$V^{imp}=1 t$. The pairing strength is set $V_{int}=-2 t$ and
chemical potential $\mu=0$ (see text). \textbf{a}, calculated local
$\frac{dI}{dV}$ tunneling conductance at positive bias $V=1.2 t$.
\textbf{b}, calculated local $\frac{dI}{dV}$ tunneling conductance
at negative bias $V=-1.2 t$. \textbf{c},  corresponding Bogoliubov
angle $\Theta(x,y)$. \textbf{d}, logarithm of the absolute value of
Fourier transform of BA $\log_{10}(|\Theta(k_x,k_y)|)$ with
subtracted average value $<\Theta(x,y)>=45^\circ$.  The Fourier
transform we obtain is consistent with FT intensity us seen in the
experiment, see inset in Fig.~\ref{fig_exp3} \textbf{e}. Please note
rotation of $(q_x,q_y)$ basis.}
\end{center}
\end{figure}

 Using the Bogoliubov-Valatin transformation to the
 quasi-particles operators $\gamma_{n\sigma}$,
\begin{eqnarray}
\label{transf} c_{i\uparrow}=\sum_n[\gamma_{n\uparrow}u_n({\bf
r}_i)-\gamma_{n\downarrow}^\dag v_n^*({\bf r}_i)],\nonumber\\
c_{i\downarrow}=\sum_n[\gamma_{n\downarrow}u_n({\bf
r}_i)-\gamma_{n\uparrow}^\dag v_n^*({\bf r}_i)],
\end{eqnarray}
and the mean-field approximation, one can diagonalize the
Hamiltonian Eq.(\ref{1}). The quasiparticle amplitudes on lattice
sites $(u_n({\bf r}_i),v_n({\bf  r}_i))$ have to satisfy
inhomogeneous Bogoliubov-de Gennes equations \cite{franz}:
\begin{eqnarray}
\label{matrix_equation} \left(\begin{array}{cc} \hat{\xi} & \hat{\Delta}\\
\hat{\Delta}^* &-\hat{\xi}^*
\end{array}\right)\left(\begin{array}{c} u_n({\bf  r}_i)\\ v_n({\bf r}_i)
\end{array}\right)=E_n\left(\begin{array}{c}  u_n({\bf  r}_i)\\ v_n({\bf r}_i)
\end{array}\right),
\end{eqnarray}
where  the kinetic operator $\hat{\xi}$ and superconducting order
parameter  $\hat{\Delta}$ can be represented as:
\begin{eqnarray}
\label{3} \hat{\xi}u_n({\bf r}_i)&=&-t \sum_{{\boldsymbol\delta}}
u_{n}({\bf r}_i+{\boldsymbol\delta}) +(V^{imp}({\bf r}_i) -\mu) u_n({\bf r}_i),\nonumber\\
\hat{\Delta} v_{n}({\bf r}_i)&=& \sum_{{\boldsymbol\delta}}
\hat{\Delta}_{{\boldsymbol\delta}}({\bf r}_i)v_{n}({\bf
r}_i+{\boldsymbol\delta}),
\end{eqnarray}
where  ${\boldsymbol\delta}=\pm{\bf \hat{x}},\pm{\bf \hat{y}}$ are
nearest neighbor vectors for a square lattice.

 We solve
Eq.(\ref{matrix_equation}) together with the self-consistency
condition:
\begin{eqnarray}
\label{4} \Delta_{\boldsymbol\delta}({\bf
r}_i)=\frac{V_{int}}{2}\sum_n(u_n({\bf
r}_i+{\boldsymbol\delta})v^*_n({\bf r}_i)+\nonumber\\u_n({\bf
r}_i)v^*_n({\bf r}_i+{\boldsymbol\delta}))\tanh(E_n/2k_BT),
\end{eqnarray}
where the summation is over the positive eigenvalues $E_n$ only.

For a square lattice system with $L\times L$ lattice sites, the
solution of  the Bogoliubov-de Genes  equations Eq.(\ref{3}) is
equivalent to the eigenproblem for a $2L^2\times 2L^2$ matrix. In
order to minimize the boundary effects we assume periodic boundary
conditions in both $x$ and $y$ directions.

We have performed numerical simulations on a square lattice
$32\times 32$ at $T=0$. We assume $40$ impurities randomly placed on
the lattice, each impurity has strength $V^{imp}=1 t$. It
corresponds to approximately $3\%$ doping. We set $V_{int}=-2t$ and
a half-filled band $\mu=0$.

The results of our numerical simulations are summarized in
Figs.~\ref{fig_theor1},\ref{fig_theor2},\ref{fig_theor3}, where we
show three panels of plots for calculated  local tunneling
conductance  $dI/dV$ at positive and negative bias, the
corresponding Bogoliubov angle $\Theta(x,y)$, and the logarithm of
the absolute value of its Fourier transform. We consider the
following values for the bias: $V=\pm 0.4t, \pm 0.8t, \pm 1.2 t$.
Note, that at bias $V=\pm 0.4t$, that is under the gap value
$\Delta\approx 0.8t$,  the pattern of the local Bogoliubov  angle
(see Fig. \ref{fig_theor1}\textbf{c} is rotated $45$ degrees with
respect to the patterns calculated at higher biases $V=\pm 0.8t$
(near the gap, see Fig.~\ref{fig_theor2}\textbf{c}, and $V=\pm 1.2
t$ (above the gap, see Fig.~\ref{fig_theor3}\textbf{c}). The sites
on the lattice where there is a large particle-like component of the
Bogoliubov excitation, hole component is small. Complementary
pattern is observed on opposite bias. This ``rotation'' is commonly
present in the whole field of view.

\begin{figure}[htb]
\begin{center}
\includegraphics[width=6cm]{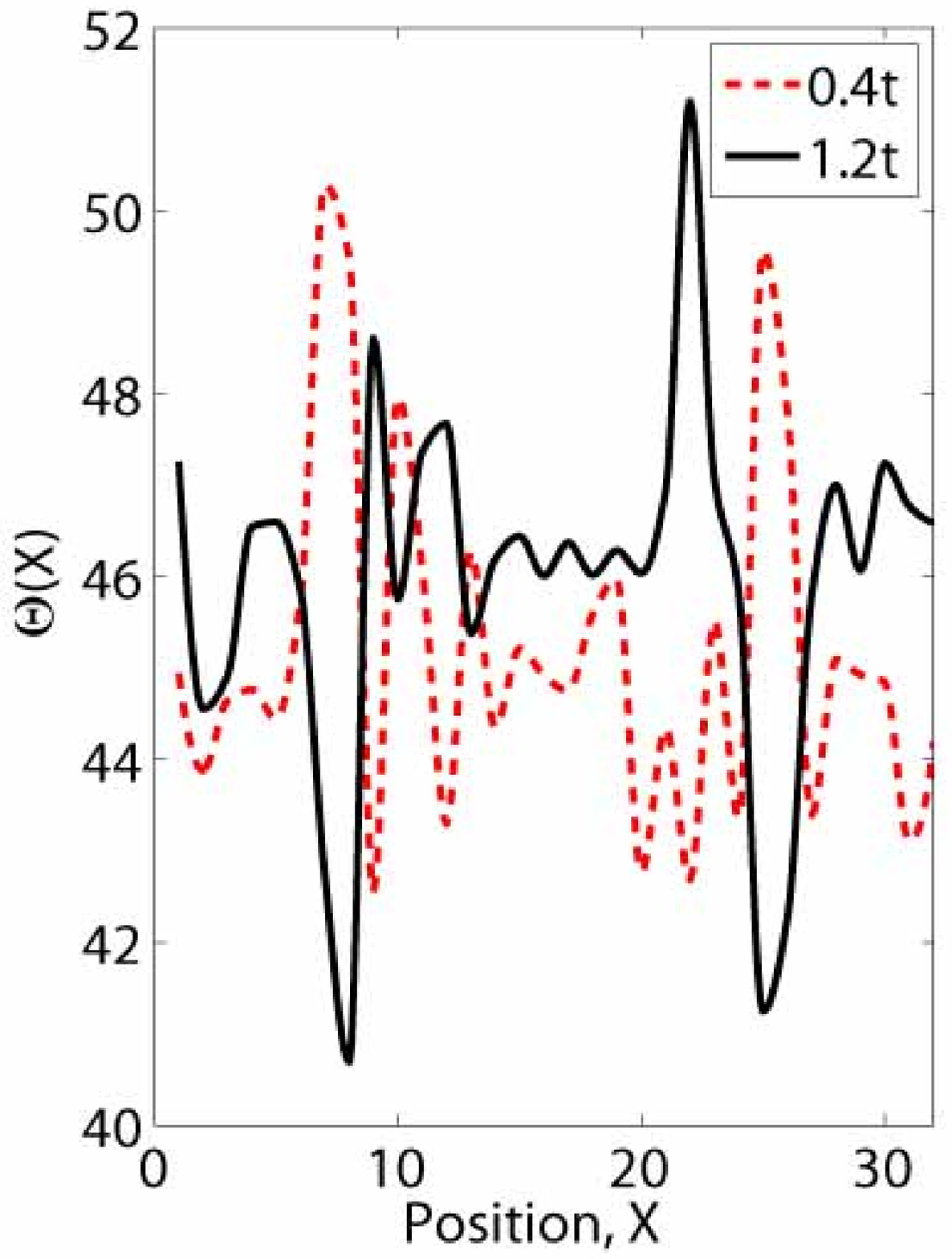}
\caption{ \label{fig_theor4}(Color online) Profile of the Bogoliubov
angle $\Theta$ along the line cut for bias values $0.4$t, red
(dashed) line; and $1.2$t, black (solid) line. The line cut is taken
along the direction [1,1]. Note the angle inversion effect with
respect to the optimal mixing angle value of $45^\circ$ for low
energy and high energy BA. }
\end{center}
\end{figure}

We also present BA along the diagonal line cut for our numerical
calculation to compare  with experimental results. We observe an
out-of-phase angle change for low energy vs high energy BA, Fig.
\ref{fig_theor4}. This out of phase behavior is consistent with the
behavior seen in Fig.\ref{fig_exp2}.

\section{Conclusion}


In conclusion, we have introduced a new spectroscopic measure,
Bogoliubov angle $\Theta(\br_i,E)$. This measure allows one to
image local particle-hole admixture in the superconducting state
and in the normal state with superconducting correlations.

Bogoliubov angle can be studied as a function of position. It can
also contain nontrivial  Fourier components.   This could allow us
to make connection with the spatial interference of quasiparticles
in superconducting state. \cite{Wang,Capriotti,Zhu,Hanaguri}
Complementary to the momentum space information one can look at the
energy dependence of BA. Energy dependence observed experimentally
clearly indicates that there is a change in behavior in
$\Theta(\br,E)$ at $E \simeq 20-25 mV$,
Figs.\ref{fig_exp2},\ref{fig_exp3}.  This energy range clearly
correlates with the changes in the interference patterns. We
interpret these changes as an evidence for a change in the
superconducting coherence that is weakened at higher energies.

As a future application we propose that BA  be studied as a function
of doping and temperature. One can use to investigate BA and
particle-hole  at temperatures above $T_c$ for studies of the nature
of the pseudogap phase. Using Bogoliubov angle one could be able to
identify how robust the particle-hole mixture is in the normal state
and therefore be able to differentiate between different scenarios
of PG state.

We are grateful to I. Martin,  A. Yazdani, N. Nagaosa, M. Randeria
and O. Fischer for enlightening discussions. This work was supported
by the BES and LDRD funds from U.S. Dept. of Energy at Los Alamos
National Laboratory under Contract No. DE-AC52-06NA25396.
 J.C.D acknowledges support from Brookhaven National Laboratory
under Contract No. DE-AC02-98CH1886 with the U.S. Department of
Energy, from the U.S. Department of Energy Award DE-FG02-06ER46306,
and from the U.S. Office of Naval Research.

\setcounter{equation}{0}
\renewcommand{\theequation}{A-\arabic{equation}}
\section*{APPENDIX A: Anderson Mapping}


Here we recall the Anderson \cite{pwa58} mapping of reduced BCS
model on the effective spin model. The reduced BCS Hamiltonian is
taken to be

\beqa H_{red} = -\sum_{\bk} (\epsilon_{\bk} - \mu) (1 - n_{\bk} -
n_{-\bk}) - \nonumber\\
 \sum_{\bk \neq \bk'}
V_{\bk,\bk'}c^\dag_{\bk} c^\dag_{-\bk}
c_{-\bk'}c_{\bk'} = \\
-2 \sum_{\bk} (\epsilon_{\bk} - \mu) s_{z, \bk} - 1/2 \sum_{\bk,
\bk'} ( s^+_{\bk}s^-_{\bk'} + s^+_{\bk'}s^-{\bk})
\label{EQ:Mapping1} \eeqa where we assumed translational
invariance for simplicity and omit spin indexes. Spin operators
are defined as: \beqa s_{z,\bk} = 1- n_{\bk} - n_{\bk'}, \\
s^+_{\bk} = b^+_{\bk} = c^\dag_{\bk} c^\dag_{-\bk} , \\ s^-_{\bk}
= b_{\bk} = c_{\bk} c_{-\bk} \label{EQ:mapping2} \eeqa and they
represent a complete spin algebra over space $n_{\bk}- n_{-\bk} =
0$, the so called hard core boson constraint. z component of the
spin corresponds to state with well defined particle number and
$s^\pm$ corresponds to pairing correlations. Anderson showed that
this reduced Hamiltonian describes the spin $s_{\bk}$ in an
``external'' field   pointing at an angle $\Theta_{\bk}$ \beqa
 \Theta_{\bk} = 1/2 \frac{\sum_{\bk'} V_{\bk \bk'} \sin
 \Theta_{\bk}}{\epsilon_{\bk} - \mu}
 \label{EQ:mapping3}
 \eeqa
 One immediately recognizes this a self-consistency equation for BCS solution,
  once we assume $V_{\bk \bk'}$ to be constant in a range near Fermi surface.
   Excitation spectrum for the effective spin model is:
 \beqa
 E_{\bk} = ( (\epsilon_{\bk} - \mu)^2 + 1/4(\sum_{\bk'} V_{\bk \bk'}
 \sin\Theta_{\bk'})^2)^{1/2}
 \label{EQ:mapping4}
 \eeqa
 Complete identification with Bogoliubov quasiparticles is clear if
 one identifies
 \beqa
 \sin \Theta_{\bk} = 2 u_{\bk}v_{\bk}, \cos \Theta_{\bk} = u^2_{\bk}
 - v^2_{\bk}
 \label{EQ:mapping 5}
 \eeqa

 To make a contact with BA we notice that in
 case of  broken translational symmetry we can work out exactly the
 same representation based on eigenfunctions  in real space $u_{n}(\br_i),
 v_{n}(\br_i)$. Then the mapping on the spin problem will be done in
 real space, angle $\Theta_{\bk}$ is proportional to the
 BA defined in the Introduction, and we will have $\Theta_{E}(\br_i)$ as defined in
 Eq.(\ref{EQ:SCangle}). One can immediately see the direct
 connection with the Anderson angle used in this effective spin
 model.

\setcounter{equation}{0}
\renewcommand{\theequation}{B-\arabic{equation}}
 \section*{Appendix B: Numerical details }

The numerical solution of Eq.(\ref{3}) together with the
self-consistency condition Eq.(\ref{4}) requires iterative solution
and  it is organized as follows: \\(1) For a reasonable initial
value of the order parameter $\Delta_{\boldsymbol\delta}({\bf r}_i)$
we solve the eigenproblem Eq.(\ref{3})  to obtain the quasiparticle
amplitudes $(u_n({\bf r}_i),v_n({\bf  r}_i))$ and the quasiparticle
spectrum $E_n$.
\\ (2) Then, substituting
$(u_n({\bf  r}_i),v_n({\bf  r}_i))$ and $E_n$ into Eq.(\ref{4}) we
compute a new approximation of the  order parameter
$\Delta^{(appr)}_{\boldsymbol\delta}({\bf  r}_i)$.\\ (3) In order to
avoid numerical instabilities during iterations, we use a mixing
scheme
  $\Delta^{(n+1)}_{\boldsymbol\delta}({\bf  r}_i)= \alpha \Delta^{(appr)}_{\boldsymbol\delta}({\bf  r}_i)+
  (1-\alpha)\Delta^{(n)}_{\boldsymbol\delta}({\bf  r}_i)$, where $\Delta^{(n)}_{\boldsymbol\delta}({\bf  r}_i)$
   is the  order parameter at  the previous iteration step. Adjustable parameter $\alpha$ is
  a number between 0 and 1. To insure convergence, we increase the current value of
  $\alpha$ by $5\%$ if the relative deviation between two consequent steps
   $S^n=\max_{i,{\boldsymbol\delta}}|\Delta^n_{\boldsymbol\delta}({\bf  r}_i)- \Delta^{n-1}_{\boldsymbol\delta}({\bf  r}_i)|/
   \max_{i,{\boldsymbol\delta}}|\Delta^n_{\boldsymbol\delta}({\bf  r}_i)|$ has
   decreased,  $S^n<S^{n-1}$. And we decrease $\alpha$ by $20\%$ in the opposite
   case, $S^n>S^{n-1}$.\\
(4) The computed $\Delta^{(n+1)}_{\boldsymbol\delta}({\bf  r}_i)$ is
used for the next iteration step.

 We repeat iterations until we achieve the
acceptable level of accuracy ($\epsilon=10^{-3}$). After the end of
the procedure, we perform an additional step with $\alpha=1$ to
ensure convergence of the  obtained solution. It usually takes
$20-40$ iterations to converge.

{99}


\end{document}